\documentclass[journal]{IEEEtran}
\usepackage[edges]{forest}
\usetikzlibrary{arrows.meta}
\usepackage{graphicx,times,amsmath,cite,enumerate}
\usepackage{kantlipsum}
\usepackage{threeparttable}
\usepackage{times}
\usepackage{epsfig}
\usepackage{amssymb}
\usepackage{latexsym}
\usepackage{psfrag}
\usepackage{graphicx}
\usepackage{setspace}
\usepackage{color}
\usepackage{verbatim}
\usepackage{subfigure}
\usepackage{amsthm}
\usepackage{makecell}
\usepackage{footnote}
\usepackage{array}
\newcolumntype{P}[1]{>{\centering\arraybackslash}p{#1}}
\usepackage{float}
\usepackage{multirow}
\usepackage{textcomp}
\usepackage{forest}
\PassOptionsToPackage{hyphens}{url}
\usepackage{url}
\usepackage{physics}
\usepackage{hyperref}
\usepackage{etoolbox}
\usepackage{tikz}

\begin{document}
\title{ Public Plug-in Electric Vehicles + Grid Data: \\  Is a New Cyberattack Vector Viable?}

 \author{ Samrat~Acharya,~\IEEEmembership{Student Member,~IEEE}, Yury~Dvorkin,~\IEEEmembership{ Member,~IEEE}, Ramesh~Karri,~\IEEEmembership{Fellow,~IEEE} \\ }

\maketitle

\begin{abstract}
High-wattage demand-side appliances such as Plug-in Electric Vehicles (PEVs) are proliferating. As a result, information on the charging patterns of PEVs is becoming accessible via smartphone applications, which aggregate  real-time availability and historical usage of public PEV charging stations. Moreover, information on the power grid infrastructure and operations has become increasingly available in technical documents and  real-time dashboards of the utilities, affiliates, and the power grid operators. The research question that this study explores is: Can one combine high-wattage demand-side appliances with public information to launch cyberattacks on the power grid? To answer this question and report a proof of concept demonstration, the study scrapes data from public sources for Manhattan, NY using the electric vehicle charging station smartphone application and the power grid data circulated by the US Energy Information Administration, New York Independent System Operator, and the local utility in New York City. It then designs a novel data-driven cyberattack strategy using state-feedback based partial eigenvalue relocation, which targets frequency stability of the power grid. The study establishes that while such an attack is not possible at the current penetration level of PEVs, it will be practical once the number of PEVs increases.

\end{abstract}
\IEEEpeerreviewmaketitle

\section{Introduction}
\label{intro}

\IEEEPARstart{T}{he} US power grid is vulnerable to attacks on its cyber infrastructure because they allow an attacker to remotely manipulate  various physical assets (e.g., generation, transmission, distribution, and substation equipment). For example, the Supervisory Control and Data Acquisition (SCADA) system of the Ukraine power grid was compromised by the BlackEnergy3 trojan, which launches a Distributed Denial-of-Service (DDoS), espionage, and information erasure attack, \cite{liang20162015}. 
First, the attackers sent spear-phishing emails with  Microsoft Word and Excel documents infected with the BlackEnergy3 trojan to  employees of the Ukraine power grid company. Second, when these attachments were opened, the trojan self-installed and automatically discovered authentication credentials of the SCADA computers. Third, the discovered authentication credentials were used to create a virtual private network channel to remotely access the human-machine interface of the SCADA system and  open circuit breakers, which led to power supply disruptions for over 225,000 end-users  \cite{blackenergy}.    
To prevent such attacks, power grid operators attempt to isolate the SCADA network from external interfaces and public networks \cite{hsdl}. Even if successful, this isolation cannot cope with demand-side cyberattacks that compromise and exploit residential and commercial high-wattage appliances. These appliances are not directly observed by  power grid operators and are vulnerable to cyberattacks due to the poor security hygiene of  end-users \cite{security_hyg} or  backdoors in their complex supply chains, involving foreign manufacturers \cite{coffin2018supply}. 

Demand-side cyberattacks uniquely differ from the previously studied utility-side cyberattacks on power grids \cite{he2016cyber, xu2018petri, sun2014multilayered, ten2017impact, singh2016stealthy,mehrdad2018cyber} due to three main reasons. First, the number of demand-side attack access points is larger than for utility-side cyberattacks due to the multi-actor and complex demand-side cyberspace managed by Plug-in Electric Vehicle (PEV) users, Electric Vehicle Charging Station (EVCS), and power grid operators. Second, the high-wattage demand-side appliances such as PEVs and EVCSs are not continuously monitored by the power grid operator, thus making it hardly possible to identify the attack on these assets, when it is launched, and to apply traditional defense mechanisms (e.g., isolation of the attacked power grid area). Third, demand-side cyberattacks  can remain stealthy to the utility, even after they are launched, because malicious power alternations are difficult to distinguish from regular power demand fluctuations. These unique aspects of demand-side cyberattacks require a state-of-the-art assessment of grid-end attack vectors for securing the power grid. 

Demand-side cyberattacks are possible because many high-wattage appliances have communication and control interfaces  forming an Internet of Things (IoT). Although such power grid attacks have not been executed in practice, similar attacks have been observed in other sectors. For instance, consider the Mirai malware that infected over 600,000 IoT devices \cite{antonakakis2017understanding}.
The Mirai malware identified and accessed IoT devices with factory-set default authentication credentials and formed a network of bots (botnet). This botnet was used to launch a massive DDoS cyberattack on the Dyn Domain Name Service  provider. The attack caused hours-long service disruptions to such web-services as Airbnb, PayPal, and Twitter \cite{antonakakis2017understanding}. As a result of this attack, Dyn lost roughly 8\% of its customers \cite{mirai_finanace}.

 Recent studies \cite{soltan2018blackiot,amini2018dynamic,dvorkin2017iot} model generic  demand-side cyberattacks on the power grid. Soltan et al. \cite{soltan2018blackiot} demonstrated that the IoT-controlled Heating, Ventilation, and Air-Conditioning (HVAC) loads can cause generator and line failures, leading to local outages and system-wide blackouts, even if a small fraction of all loads is compromised (e.g., 4 bots per 1 MW of demand, where  1 bot is considered as 1 IoT-controlled HVAC unit). Additionally, results in \cite{soltan2018blackiot} illustrated that the compromised loads can increase the operating cost (e.g., 50 bots per 1 MW demand can increase the power grid operating cost by 20\%). 
 Amini et al. \cite{amini2018dynamic} used load-altering demand-side attacks to cause power grid frequency instability over multiple periods aided by real-time frequency feedback. As \cite{amini2018dynamic} shows, multi-period attacks require a smaller number of compromised loads, relative to the single-period attacks \cite{amini2018dynamic}. 
  Dvorkin and Garg \cite{dvorkin2017iot} demonstrated propagation of demand-side cyber attacks from the distribution network to the transmission network, which scales attack impacts across large geographical areas. However, \cite{soltan2018blackiot,amini2018dynamic,dvorkin2017iot} consider generic appliances and do not consider specific attack vectors caused by a particular high-wattage, IoT-enabled appliance. Furthermore, these studies use generic power grid test beds customized for the needs of their case studies. These assumptions lead to a conservative assessment of impacts that demand-side cyberattacks have on the power grid, which can be launched by a perfectly omniscient attacker. In practice, the attacker has limited knowledge of the power grid and the compromised loads, which reduces the attack severity. This paper aims to avoid  unrealistic generalizations on the attack vector and, therefore, collects and exploits publicly accessible power grid  and EVCS demand data. This study focuses on PEVs and public EVCSs among a larger pool of demand-side attack vectors, potentially including air-conditioners, boilers, residential PEVs, because public EVCSs release their demand data publicly as a part of their business model (e.g., for the convenience of PEV users), while other demand-side attack vectors cannot be aided by data releases.
  
  Our review shows that charging patterns of high-wattage PEVs in public EVCSs are reported through smartphone applications (e.g., ChargePoint). Although attempts were made to develop cyber hygiene requirements and protocols for charging PEVs  (e.g., the 2017 report by European Network for Cyber Security \cite {encs}), there is no established consensus among manufacturers, consumer advocates, utilities, as well as  national and international authorities. For instance, power utilities in New York proposed a cybersecurity protocol, which was subsequently denied by the third-party service providers due to its engineering and cost implications \cite{ny_cyber}. Most of the utilities still treat PEVs and EVCSs as passive loads and do not pro-actively monitor their usage and cyber hygiene. As a result, the cybersecurity community warns that PEVs and EVCSs can evolve as an attack vector into the power grid. For instance, Kaspersky Lab revealed security flaws in the ChargePoint Home charger and its smartphone application \cite{chpoint_cyber}. This flaw would enable an attacker to remotely control PEV charging after gaining  access to a Wi-Fi network
 
  to which the charger is connected. Fraiji et al. \cite{fraiji2018cyber}, Ahmed et al. \cite{ahmed2016electric}, and Pratt and Carroll \cite{8662043} discuss cyber vulnerabilities of communication interfaces of IoT-controlled PEVs and EVCSs. The vulnerabilities in \cite{chpoint_cyber, fraiji2018cyber, ahmed2016electric, 8662043} are considered from the viewpoint of an attack damaging either PEVs or EVCSs. However, threats imposed on the power grid from such vulnerabilities are not assessed. 
 
This paper aims to demonstrate that public information on EVCS demand and power grids is  a cyber threat to the urban power grid with a large PEV fleet. The study appraises the risk of realistic rather than omniscient attack assumptions by only using public data to represent EVCS demand and power grid operations for designing the attack. Our main contributions are summarized as follows:

 \begin{enumerate}
 \item The paper is the first of its kind to evaluate the power grid  vulnerability to an unsophisticated demand-side cyberattack strategy derived using publicly available power grid, EVCS, and PEV data. Such \textit{dilettante but realistic} vulnerability assessments are not common in power grid security analyses, which typically employ the worst-case attack assumption (e.g., an omniscient, insider attack with a perfect knowledge of the system), but common in other disciplines. Therefore, to understand the critical nature of publicly available data on the power grid, this paper adopts a design of the $N^{\text{th}}$ Country Experiment carried out by the Lawrence Livermore National Laboratory in the 1960-s, which aimed to assess the ability of non-military personnel to  design a military-grade nuclear explosive  device using publicly available materials. The team of three physicists designed such a device, i.e., a so-called `dirty bomb', within 2 years, which had long-term implications for nuclear nonproliferation \cite{arbatov2011beyond}.
 
 \item As a proof of concept, this paper  demonstrates how an attacker can collect data on PEVs, EVCSs, and the power grid using  public sources from Manhattan, NY. Using this data, the study designs a novel data-driven attack strategy that manipulates PEV and EVCS loads to cause frequency instability in the power grid. The novelty of this data-driven attack strategy is that it builds on the state-feedback-based partial eigenvalue relocation using the Bass-Gura approach, which makes it possible to relocate some eigenvalues toward the locations chosen by the attacker and to minimize the amount of the compromised demand needed for the attack. Unlike previously studied attacks based on real-time measurement and feedback, e.g., as in \cite{amini2018dynamic, murguia2018reachable,pasqualetti2019,peng2019survey}, the attack strategy in this work does not require real-time monitoring of the power grid state, i.e., it can be carried out remotely, and can be robustified to the ambiguity in estimating the EVCS demand. Thus, informed by real-life data availability,  this paper  avoids assuming an  omniscient attacker modeled in \cite{amini2018dynamic,soltan2018blackiot,dvorkin2017iot,pasqualetti2019,yuan2011modeling}.
 
\item Based on extensive numerical simulations using real-life data, this paper summarizes the identified cyber vulnerabilities that can be exploited as access points to launch data-driven, demand-side cyber attacks using grid-end attack vectors. Finally, we anticipate that this paper will raise awareness about the simplicity of designing and executing  data-driven, demand-side cyberattacks and facilitate the negotiation of a common cybersecurity protocol \cite{ny_cyber} for high-wattage appliances.
\end{enumerate}

\section{Public EVCS and Power Grid Data}
\label{data}
 \begin{figure}[b]
\includegraphics[width=1.0\columnwidth]{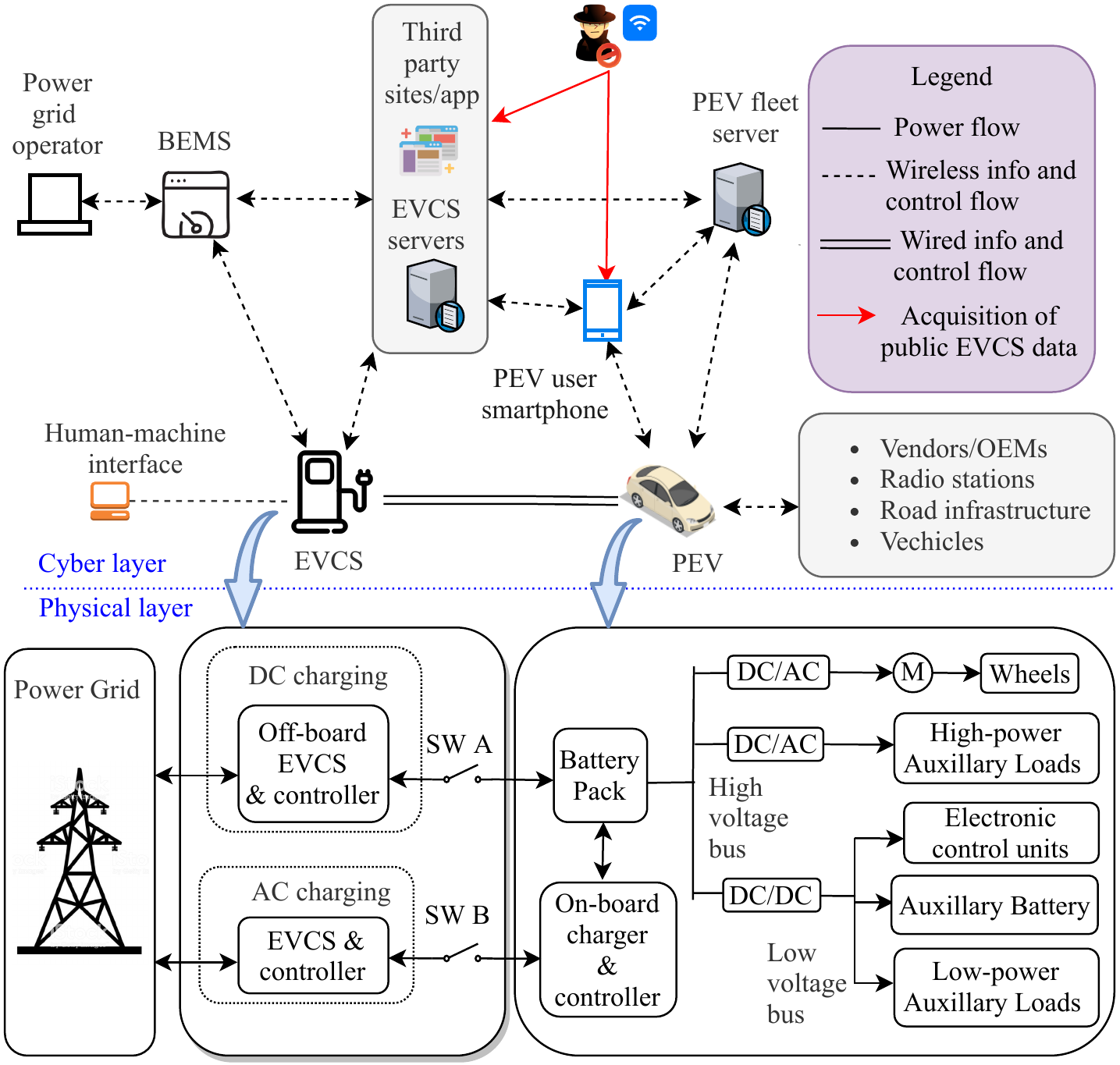}
\caption{ Cyber-physical interfaces among the power grid, EVCSs, and PEVs, as well as sources of public EVCSs data.}
\label{fig:EV1}
\end{figure}

Analyzing cybersecurity of smart grids rests on a cyber-physical model describing its physical assets, cyberinfrastructure, and their interlinks \cite{he2016cyber, xu2018petri}. This section describes cyber-physical interfaces among the power grid, EVCSs, and PEVs and details a procedure to collect public data that the attacker can use to plan and launch an attack. Since the borough of Manhattan, NY has the greatest penetration rate of PEVs in the state of New York \cite{EVdemand}, we considered this area to demonstrate the attack concept.

\subsection{Interdependence between the Power Grid and PEVs}
\label{cyber_physical}
 Fig. \ref{fig:EV1} shows the cyber and physical links between the power grid, the EVCSs, and the PEVs. An EVCS is pivotal to the cyber-physical interdependency between the power grid and PEVs. The attacker can observe some of these interdependencies using web-services of the EVCS vendors and the third-parties like ChargePoint that aggregate EVCSs and PEVs data. We describe the interdependence between the physical and cyber layers below.

\subsubsection{Physical Layer}
 A typical EVCS hosts power converters (AC/DC and DC/DC), power conditioning units (e.g., power factor corrector), sensors, and controllers that enable and direct the power flow between a power grid and PEVs. Although this power flow can be bidirectional, the flow from PEVs to the grid is not yet always commercial. As shown in Fig.~\ref{fig:EV1}, the EVCS charging circuit can be broadly split into off-board charging (DC) circuit and on-board charging (AC) circuit. Thus, in the off-board charging circuit, the EVCS converts AC power from the grid to DC to  charge the PEV battery. On the other hand, this AC/DC power conversion takes place inside the PEV in the on-board charging circuit. Although many PEVs support both on and off-board charging options, they cannot be used simultaneously due to mutually exclusive switches SW A and SW B as shown in Fig.~\ref{fig:EV1} \cite{TI}.
 
 There are three EVCS levels adopted by vendors and related organizations: Level 1 (L1), Level 2 (L2), and Level 3 (L3). They vary in their power capacity, voltage, and current ratings. Generally, the L1 and L2 EVCSs need the on-board charging circuit in PEVs, whereas the L3 EVCSs have the off-board charging circuit. The L1 EVCS are wall outlets in a residential single-phase AC system rated at 120 V, 12-16 A, and deliver 1.44-1.9 kW power to the PEVs from the grid. Since  L1 chargers are typically installed in homes, their data is not generally available to the public. The L2 and L3 chargers are in commercial charging stations that can host multiple PEVs. The L2 EVCS uses a single or split-phase AC system with 208-240 V, 15-80 A, and delivers 3.1-19.2 kW power to PEVs from the grid. The L2 EVCS charges PEVs faster than the L1 EVCS. The L3 EVCS is the most high-wattage PEV charger and, therefore, induces greater volatility to  power grid operations. These superchargers are DC systems with 300-600 V, up to 400 A, and deliver 25-350 kW to each PEV. Notably, there are various types of connectors between EVCSs and PEVs that vary depending on different geographical areas and EVCS levels.

Within a PEV, a PEV battery delivers power to a three-phase AC traction motor through a DC/AC traction inverter. Most of the PEVs use a permanent magnet synchronous motor or an induction motor (e.g., Tesla). The DC/AC traction inverter is bidirectional in some PEVs, which allows for charging the battery by regenerative braking. Auxiliary loads in a PEV, such as an air-conditioning compressor, are supplied through DC/AC inverters. The AC traction motor and the air-conditioning compressor are usually operated at a higher voltage (typically 240 V), while other low-power auxiliary loads, back-up batteries, Electronic Control Units (ECUs), and power steering operate at a lower voltage (typically 12-48 V). 
 
\subsubsection{Cyber Layer}
 The cyber interface among the power grid, EVCSs, and PEVs is intricate and constantly expands, which increases the difficulty of its generalization. Below we summarize the state-of-the-art PEV-EVCS-grid interfaces in the US, which is relevant for the attack vector considered in this paper.
 
As shown in Fig.~\ref{fig:EV1}, EVCSs communicate with PEVs via a wired communication channel used to control a PEV charging process. However, this wired communication varies for L1, L2, and L3 EVCSs. Thus, the L1 and L2 EVCSs use a pilot wire, while the  L3  EVCSs  communicate  using  either the Controller Area Network (CAN)  or the  Power  Line  Communication  (PLC)  protocol. The exchanged information between the PEV and EVCS via the wired communication channel includes the availability of a PEV for charging using Pulse Width Modulation (PWM) signals, charging current, state of charge of the PEV battery, and ground-fault detection \cite{TI}.  Generally, the  L1  EVCSs  are  simple,  private,  and  stand-alone. They  neither  communicate  with  users  nor   are they networked  with  other  EVCSs.  Therefore, the  EVCS cyber links discussed in this section  mostly  apply  to  L2  and  L3  EVCSs. The latter EVCSs communicate with their users via a Human Machine Interface (HMI), typically through an on-site card reader with a touchscreen or a smartphone application. This HMI allows PEV users to customize their charging session, i.e., to select the type of the EVCS connector, method of payment (cash or card), charging  duration, charging power rate, etc.  Moreover,  the  HMI displays a charging  price  and an EVCS operating status in real-time.  
 
Commercial L2 and L3 EVCSs are networked via a Wide  Area  Network (WAN) to enable their centralized control and to interface with the power grid. The basic function of the centralized EVCS server (see Fig.~\ref{fig:EV1}) is to collect charging session information from EVCSs, authenticate and authorize a PEV user to charge, inform the PEV users about EVCS availability, and schedule interactions with the power grid. PEV users interact with an EVCS server via smartphone applications. The EVCS server may further coordinate with the Building Energy Management System (BEMS) to control the participation of EVCSs in Demand Response (DR) events \cite{encs}. In current practices, the BEMS receives DR calls from a power grid operator or a DR aggregator, and controls the EVCS power consumption via a Home Area Network (HAN).

 On average, a typical PEV has more than 70 ECUs connected with the Controller Area Network (CAN) bus standard. The ECUs consist of microprocessors, memory units, and input/output interfaces. Some examples of ECUs include an engine control module, a battery management unit, and an air-conditioner unit. The CAN bus architecture is based on a peer-to-peer network, where each ECU and peripheral units are peers. The CAN bus standard is adopted in PEVs due to its capability to handle simultaneous commands from multiple ECUs in real-time and without communication delays. Further, the CAN bus is flexible for adding and removing ECUs and is cost-effective and robust towards electric disturbances and electromagnetic interference \cite{CAN}. However, the CAN bus standard is designed based on an isolated trust model and, therefore, does not account for the security threats from external communications (e.g., see the discussion in Section~\ref{vul_PEV}).

PEVs communicate to external networks via wired, e.g., Universal Serial Bus (USB) ports, Compact Disks (CDs) or Secure Digital (SD) cards, or wireless technologies, e.g., WiFi, Bluetooth, Near-Field Communication (NFC), Radio Frequency (RF), and cellular networks. For example, the ECUs and the infotainment system have  external communication to their vendors or Original Equipment Manufacturers (OEMs) and radio stations, respectively, via cellular networks \cite{ECU_cellular}. Further, the smart features of PEVs, such as keyless door entry, engine start, and tire pressure monitoring systems have wireless external communications. Also, PEV users connect their smartphone to the PEV via USB ports for monitoring the PEV battery charge, charging a smartphone, and accessing the smartphone via PEV dashboard. Modern PEVs also allow for wireless communications with other vehicles and road infrastructures (e.g., traffic signals) for improving driving comfort and safety.

 \subsection{Acquisition of Publicly Accessible Data}
 \label{EVCS_data}
 \subsubsection{EVCS Data}
 
 \begin{figure}[!b]
\centering
\vspace{-0.5cm}
\includegraphics[width=1.0\columnwidth, clip =true, trim = 10mm 0mm 10mm 5mm]{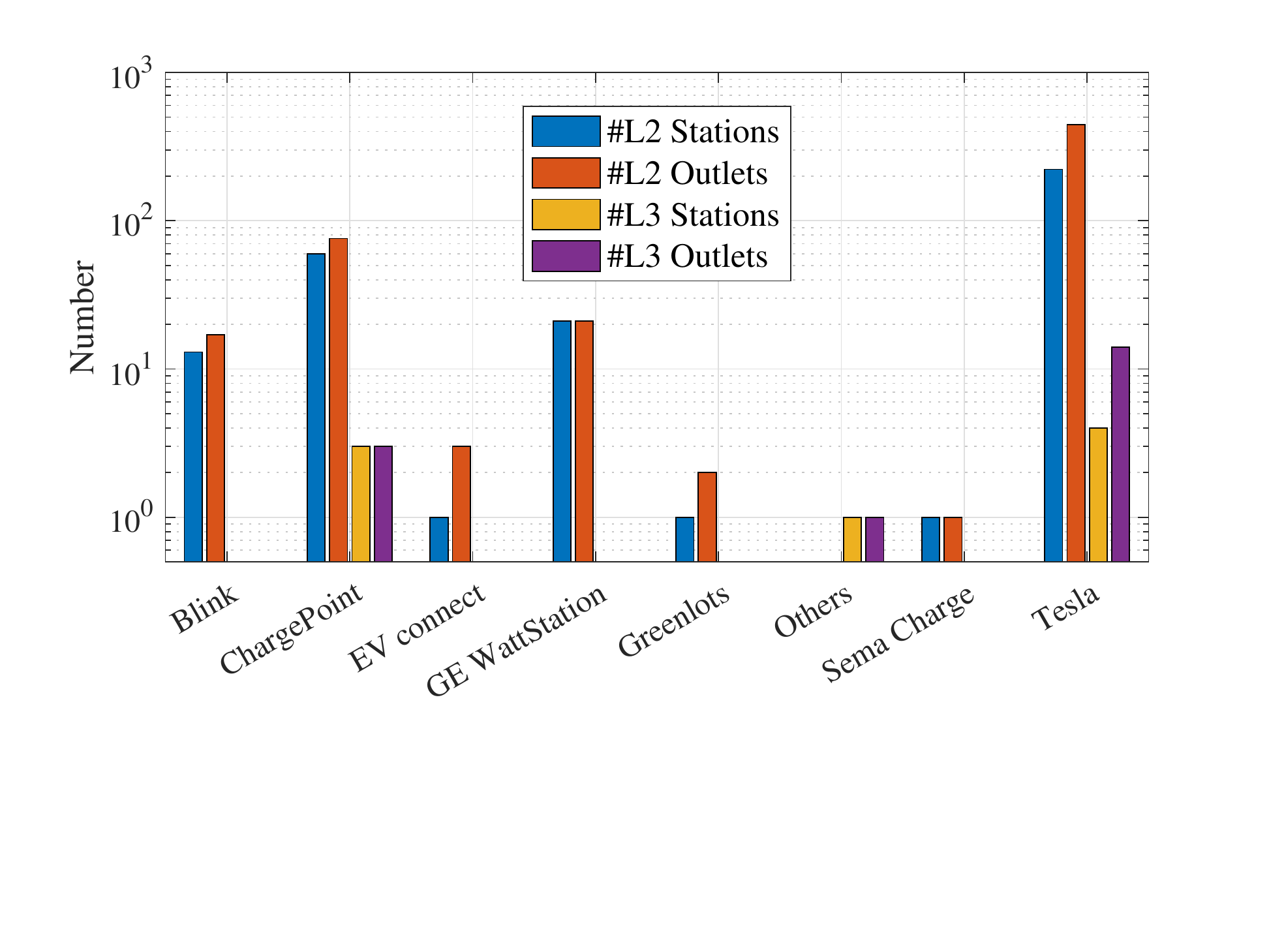}
\vspace{-2.5cm}
\caption{Number of public L2 and L3 EVCSs and their outlets (in logarithmic scale) for different EVCS companies in Manhattan, NY as of March 2019.}
\label{fig:EVCS_distribution}
\end{figure}

Commercial EVCSs are controlled by EVCS servers. The centralized EVCS server authorizes PEV users to charge their vehicle at a given EVCS, monitor and control the charging session, collect and store the session information, and release the information to users and public via smartphone applications and websites. Although EVCS companies typically have their own dedicated servers, some companies manage cross-company EVCSs. Due to this intricate nature of EVCS control and data logging, a large number of non-EVCS companies also take data from EVCS servers and release the EVCSs data publicly. These entities are referred as third-party sites/apps in Fig. \ref{fig:EV1}. Below we present how we aggregated such EVCS public data in Manhattan, NY.

\begin{figure}[!t]
\centering
\includegraphics[ width=1.0\columnwidth, clip =true,trim = 10mm 9mm 0mm 0mm]{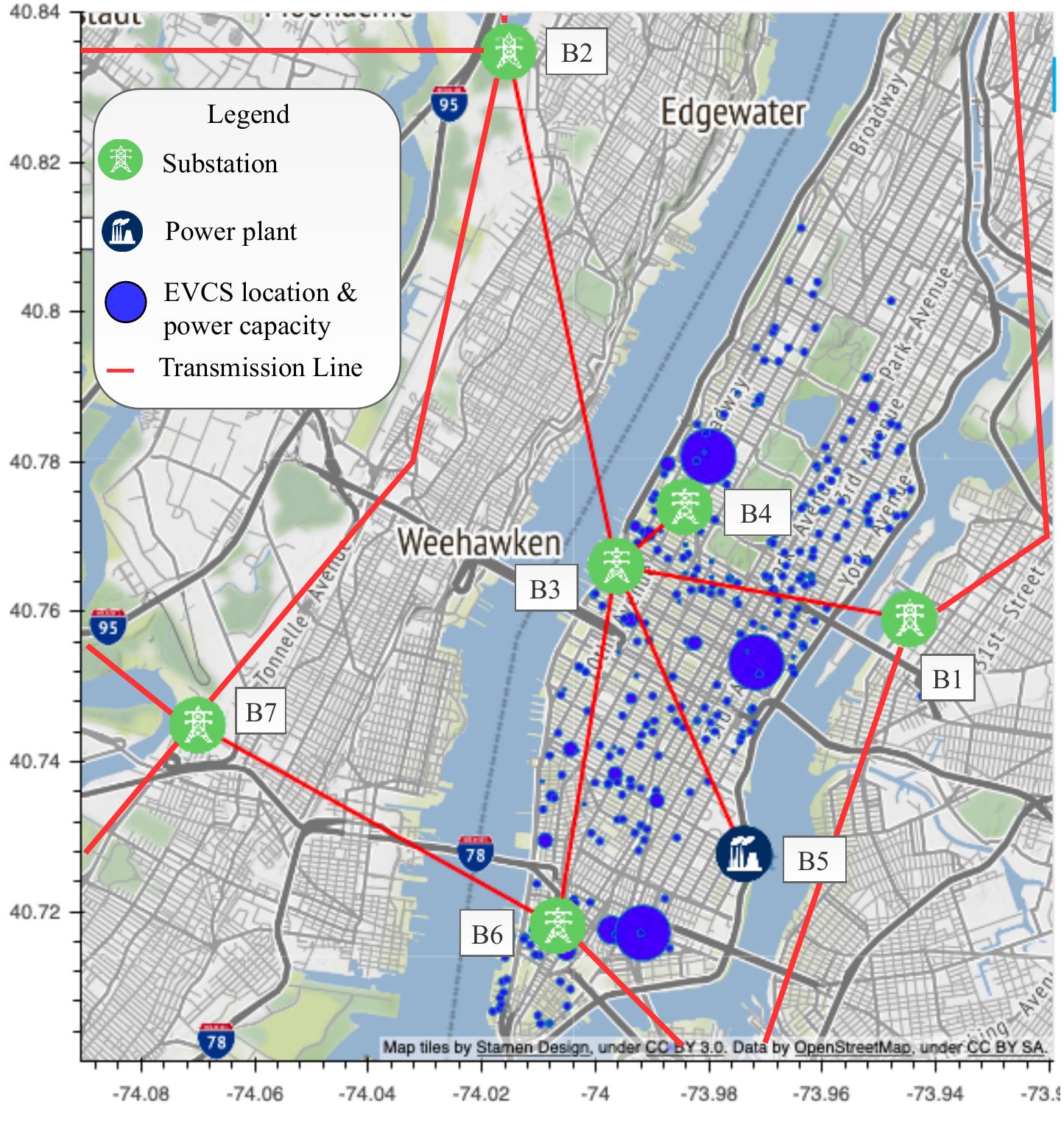}
\caption{Topology of the transmission-level power grid and locations of public EVCSs in Manhattan, NY as of March 2019. The power grid configuration includes transmission lines (138 kV and 345 kV), substations and power plants. The size of the blue circles is proportional to the EVCS demand.}
\label{fig:network_man}
\end{figure}

\begin{figure}
\centering
\includegraphics[width=1.05\columnwidth, clip =true,trim =  10mm 0mm 10mm 0mm]{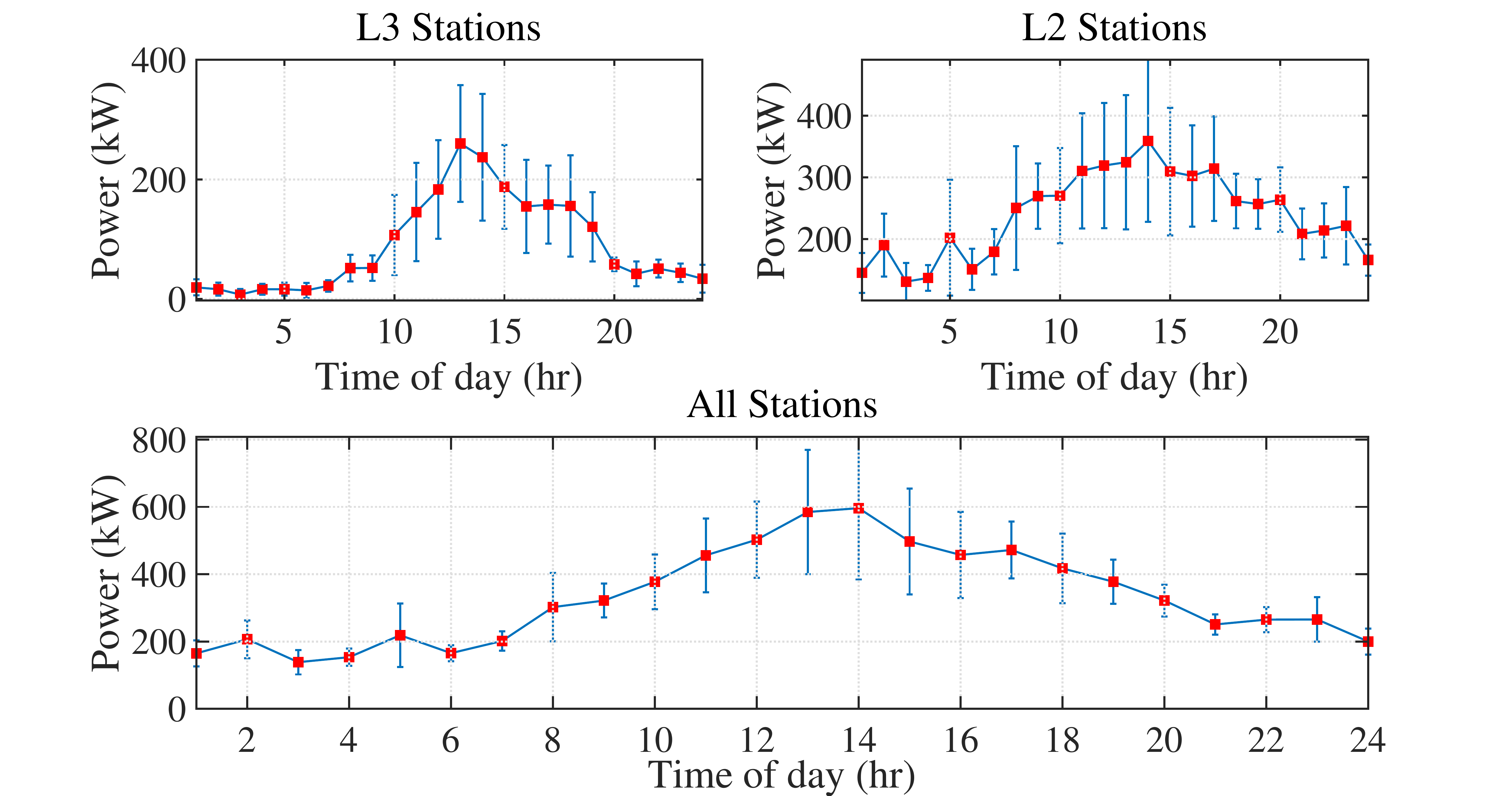}
\caption{Average hourly power consumed by public EVCSs and their standard deviation over a week in Manhattan, NY.}
\label{fig:weekall}
\end{figure}

To acquire publicly available granular information of EVCSs, we used the ChargePoint smartphone application. This application aggregates 319 L2 and L3 EVCSs operated by different companies across Manhattan, NY, as of March 2019, as shown in Fig. \ref{fig:EVCS_distribution}. Also, the Alternative Fuels Data Center (AFDC) provides information about the location and business hours of EVCSs located in the US and Canada \cite{afdc}. Cross-verifying the information provided by ChargePoint and AFDC, we obtained information about locations, power ratings, real-time and historical hourly usage profile of EVCSs as summarized in Fig. \ref{fig:network_man}. Each L2 charger in Fig.~\ref{fig:network_man} has a power rating of 6.6 kW, while the power ratings of L3 chargers are 25 and 72 kW (72 kW for Tesla superchargers). Fig.~\ref{fig:weekall} displays the total average hourly power consumed by the EVCS of each type and their standard deviation.
 
\begin{table}
\caption{Sources of publicly available power grid data for reconstructing grid parameters.}
\label{tab:info_attack}
	\vspace*{-0.1cm}
	\begin{threeparttable}[t]
	\centering
	\resizebox{\linewidth}{!}{
	\begin{tabular}{|p{2cm}|p{2.8cm}|p{4.5cm}|}
		\hline
		
		\rule{0pt}{1ex}{Information}&{Details}&{Resources}\\\hline 
		\rule{0pt}{1ex}Network Configuration &Location, capacity generators (MVA, kV), transmission lines (kV), substations (kV) & Documents from utilities, affiliates, organizations, research and development projects, e.g., \cite{eia_map,par_flow}. \\ \hline
		\rule{0pt}{1ex}Transformer, line parameters & MVA, Impedance, X/R ratio& MVA\tnote{*}, IEEE test standards, reference designs, e.g., \cite{pjm_design,transmission1964distribution}.  \\ \hline
		\rule{0pt}{1ex}Generator \newline parameters&Moment of inertia, damping coefficients& IEEE test standards, reference designs,  utilities/manufacturers catalogue, e.g., \cite{pjm_design,transmission1964distribution}.   \\ \hline
		\rule{0pt}{1ex}Generator controller \newline {parameters}&Controller type and parameters&Manual estimation referring to the white papers and scientific research papers.  
		\\ \hline
		\rule{0pt}{1ex}Load \newline parameters&Base and manipulatable loads, load damping constant& Real-time loads and historical profiles published by operator; Manipulatable loads described in Section \ref{EVCS_data}; 
Damping constant:1-2\% \cite{kundur1994power}.\\ \hline
	\end{tabular}
	}
\begin{tablenotes}\footnotesize
\item[*] Estimated from line flows associated with the substation.
\end{tablenotes}
\end{threeparttable}
\end{table}
\subsubsection{Power grid data}
\label{power_data}
Unlike the EVCS data, which is publicly available via third parties and dedicated aggregators, power grid data is fragmented. Therefore, an attacker has to manually review a vast number of documents from multiple public sources to reconstruct the grid topology and model the physical and electrical characteristics of the components.

It is noteworthy that the availability of power grid data is not uniform across power grids. This data availability correlates with the existence of markets and modernization efforts in the power grid. For instance, market-operated power grids with demand response programs publish real-time data on demand, generation, flow, and price on their websites (e.g., \cite{NYISO}). Furthermore, grid topology can be mined online regardless of the market existence. In addition to high-level locational information about power grids available through public Geographical Information System (e.g., Google Maps),  one can refine this representation of the power network using publicly reported updates on projects performed by the power utilities, which  provide information about the locations of substations, transmission lines and power plants, and specific parameters (e.g., power and voltage ratings of lines and substations). The remaining power grid parameters that the attacker needs to launch a demand-side attack are not readily available, but can be inferred from mandatory IEEE and IEC standards.

\begin{figure}[t]
\centering
 \includegraphics[width=1.0\columnwidth, clip =true,trim = 0mm 0mm 0mm 0mm]{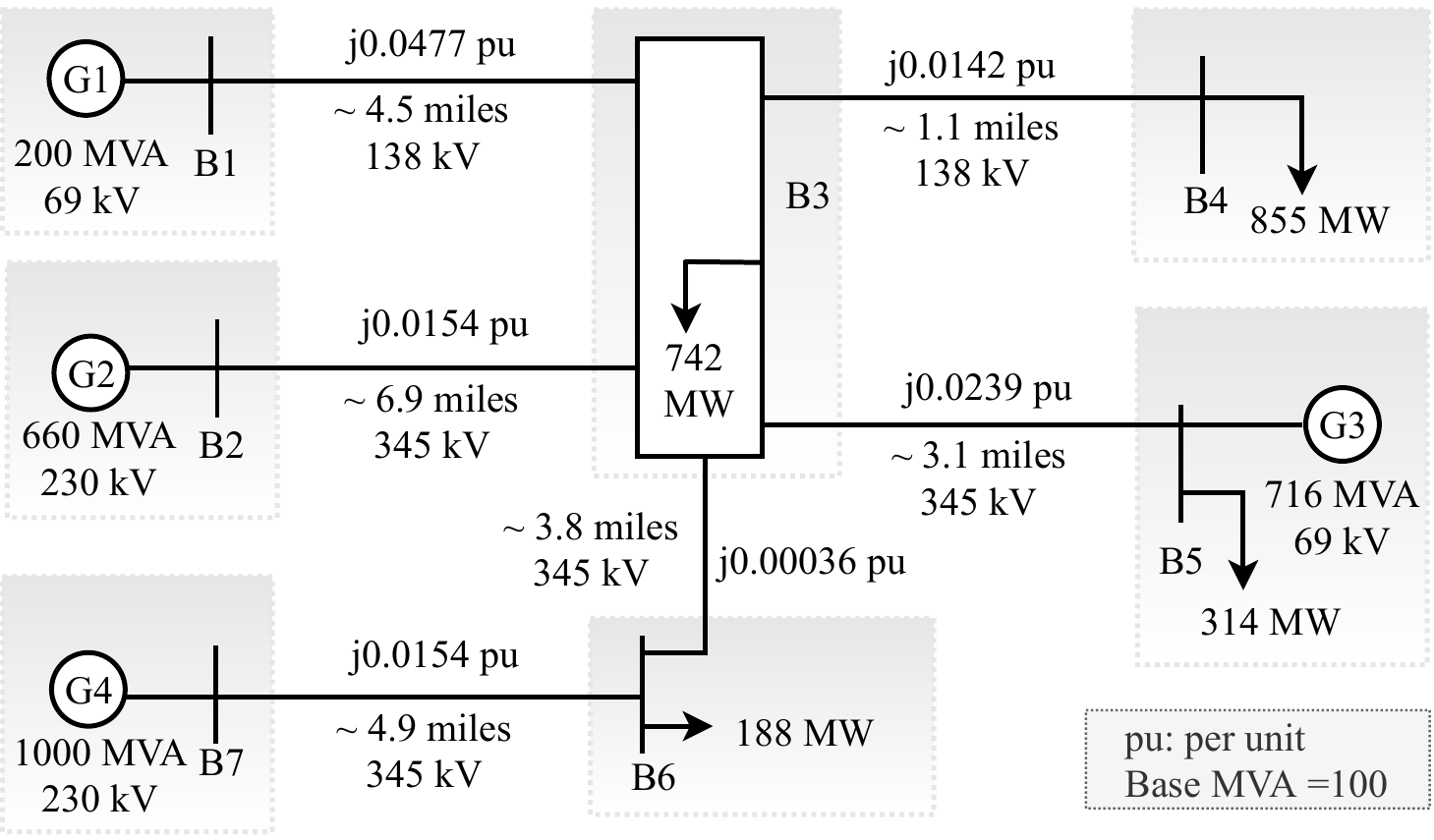}
\caption{An electrical diagram of the power grid  in Fig.~\ref{fig:network_man}. Only  transmission lines that are within Manhattan, NY and those which directly supply power to Manhattan are considered. The line lengths are estimated via Google Maps.}
\label{fig:network}
\end{figure}

In this paper, the location of generators, transmission lines, transmission substations, and their capacity (kV and MVA) are extracted from the US Energy Information Administration \cite {eia_map}. The obtained topology is cross-verified with documents released by the utilities, e.g., \cite{par_flow}. Historical and real-time generation, demand, and line flow data is learned using the real-time dashboard of the New York Independent System Operator (NYISO) \cite{NYISO}. Compiling and using this information, we were able to reconstruct a 345 kV and 138 kV transmission network configuration with  substations, lines,  power plants, and an aggregated nodal demand in Manhattan, NY. This information is shown atop EVCS location map in Fig.~\ref{fig:network_man}. Using the representation in Fig. \ref{fig:network_man}, we design an equivalent electric circuit given in Fig. \ref{fig:network}. Table~\ref{tab:info_attack} summarizes the sources of power grid data and methods to obtain the grid parameters.

The only large power plant (nominal capacity  of 716 MW) is in node B5. We model other substations as either generators or loads based on their power injection (e.g., tie lines) or consumption. The demand data for  New York City is reported by the NYISO. We itemize this demand for each node in Fig.~\ref{fig:network} using load distribution from \cite{howard2012spatial}. The power flows in tie-lines connecting the Manhattan network to the PJM Interconnection and to the rest of the NYISO system are learned from the values released in the real-time dashboard of the system operators and the utilities \cite{nyiso_def}. Impedance of the underground transmission line cables are computed using Carson’s equations \cite{kersting2006distribution} and the cable parameters in \cite{cable}. We approximated the voltage ratio of substation transformers based on the voltage level of the cables and the generating stations. Similarly, MVA-ratings of substations are approximated based on the associated generation and load. We approximated other parameters such as the transformer impedance ratio and the moment of inertia of the power plant using  data sheets for equipment with comparable parameters \cite{pjm_design,transmission1964distribution}.

\section{Attack Development}
\label{attack_seq}
\subsection{Attack Preparation}
\label{attack_prep}
\begin{figure}
\centering
 \includegraphics[width=1.02\columnwidth, clip =true,trim = 0mm 0mm 0mm 0mm]{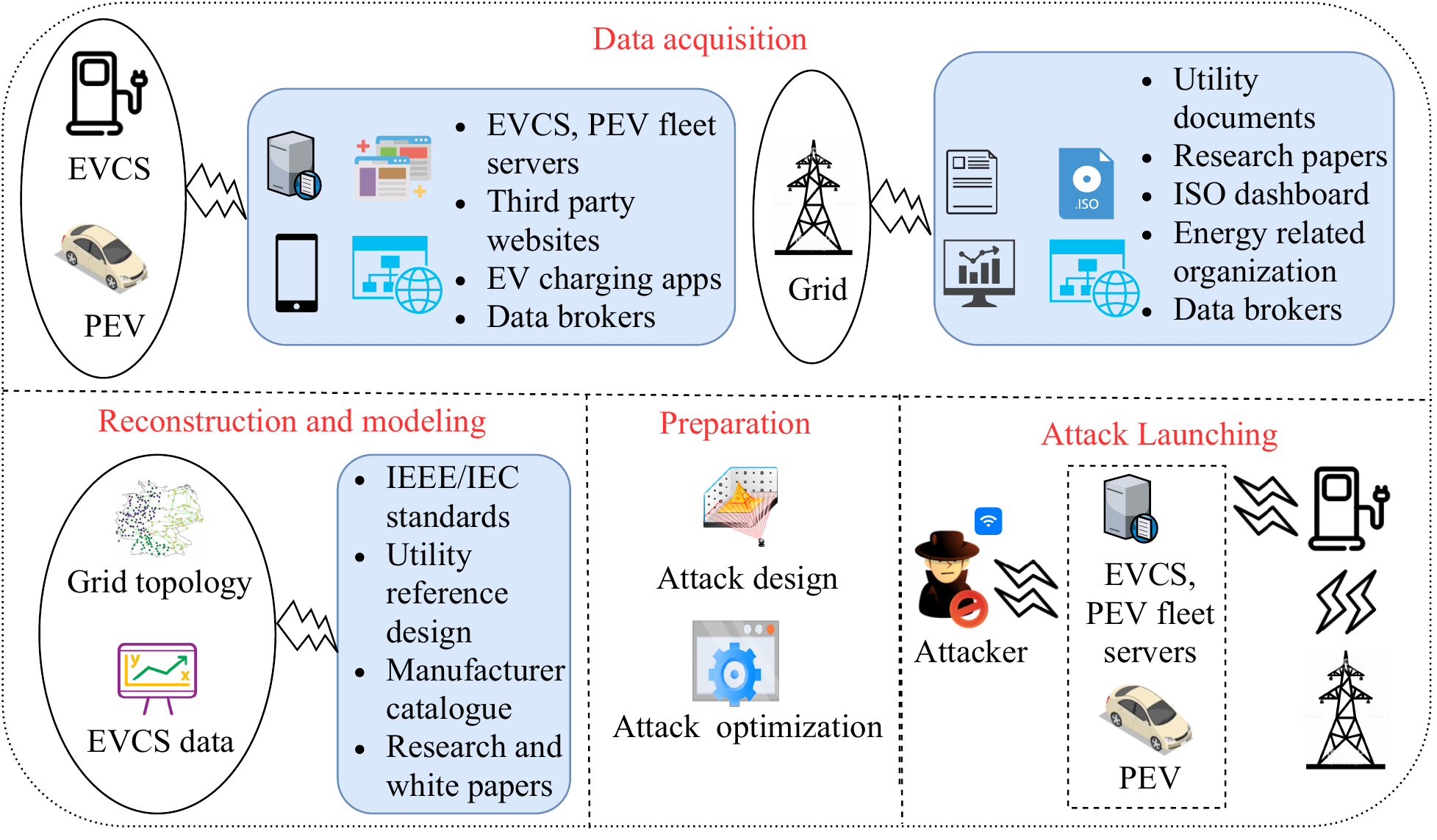}
\caption{A stage-by-stage procedure to develop a data-driven, demand-side cyberattack on power grids.}
\label{fig:attack_sequence}

\end{figure}

In this section, we describe a stage-by-stage attack development, as shown in Fig.~\ref{fig:attack_sequence}, using the cyber-physical infrastructure described in Section \ref{cyber_physical}. In the data acquisition stage, the attacker acquires the EVCS and power grid public data using the methodology described in Section \ref{EVCS_data}. Besides the publicly available power grid data, there is a growing number of data brokers (e.g., \cite{rex}), which provide granular power grid information that can be used to infer missing data. In the reconstruction and modeling stage, the attacker reconstructs the power grid configuration and the time and location-specific EVCS demand as shown in Fig.~\ref{fig:network_man} and Fig.~\ref{fig:weekall}, respectively. Then the attacker can model the reconstructed power grid parameter using the open-source power grid reference designs, IEEE standards, and research papers as explained in Table \ref{tab:info_attack}. Based on the data availability, an attacker can select  appropriate power grid cyberattack models. In the preparation stage, the attacker designs a cyberattack and optimizes the attack process. For example, the attacker may elect to optimize the attack exploiting openly-accessible information such as traffic and vehicle mobility data, weather  data, and specific social events, with the goal of increasing the efficiency of the chosen attack vector. Finally, the attacker exploits the identified vulnerabilities in the attack vector of interest and launches the attack. The discovered vulnerabilities, together with the attack development stages described above, can be structured as an attack tree or a multi-layered attack impact evaluation model \cite{sun2014multilayered} to rank all available vulnerable paths in terms of the  attack success. However, assessing all vulnerable paths is out of scope of this paper and we  outline the discovered vulnerabilities for the  attack vectors of interest (i.e., EVCS and PEVs).

\subsection{Vulnerabilities in PEVs}
\label{vul_PEV}
From the cybersecurity viewpoint, PEVs can be broken down in three components: i)  ECUs and peripherals connected via the CAN bus, ii) internet service portals such as smartphone apps and websites, and iii) communication links, such as WiFi, Bluetooth, and cellular networks, between the PEVs and internet service portals. Here, we categorize the vulnerabilities of PEVs into the internal and external, which are relevant to provide a conceptual background to the demand-side cyberattack analyzed in this paper.

\subsubsection{Internal Vulnerabilities}
Accessing the ECUs  makes it possible to fully control a given PEV. Although there are no direct paths to access the ECUs, attackers can infiltrate them either by exploiting the peripheral devices and CAN bus or by compromising  external entities (e.g., manufacturers, EVCS, BEMS) that communicate with the PEV. 

The most vulnerable ECU peripheral device is an On-board Diagnostic (OBD2) port, \cite{woo2014practical}, which is typically located under the PEV dashboard, is a standardized interface to the CAN bus that can be used by a PEV mechanic, a PEV user, and PEV regulatory authorities to monitor and obtain reports of an operational status of PEV. Since the OBD2 port is connected via the CAN bus, which is not designed to be cyber resilient\footnote{The ECU messages transported via the CAN bus are neither encrypted nor authenticated so as to reduce the memory overhead and achieve a speedy message transfer\cite{currie2017hacking}. The messages do not have address of the sending and receiving ECU. Therefore, the messages are received by all ECUs and the targeted ECU accept the messages based on the arbitration ID \cite{NI_CAN}.}, the attacker accessing the CAN bus via the OBD2 port can launch fatal cyberattacks on PEVs \footnote{The OBD2 interface is mandatory by law in the United States and Europe, and is recommended by Society of Automotive Engineers (SAE).} (e.g., DDoS attack on the ECU controlling the brakes of PEV). The OBD2 scanner is also paired with service portals and smartphone apps, which  can in turn be used for malicious intrusion into the CAN bus and ECUs remotely \cite{woo2014practical}.

Besides the OBD2 port, the attackers can also exploit USB ports, SD card ports, and CD-ROM/DVD-ROM as access points to get into the in-vehicular network. The devices that plug into these ports can be malicious or infected with malware. Moreover, the attacker can access these peripheral ports physically (e.g., a mechanic, a renter, etc.) and infect the ECUs \cite{koscher2010experimental}. Furthermore, PEVs can get infected at various points of the supply chain and maintenance. For instance, a vehicle part from an OEM or a third-party can have embeded unnoticed worms. Similarly, a malicious USB drive or SD card can furtively infect the PEV. Although these physical access points are less likely to be used for demand-side cyber attacks because they require physical access of the attacker to PEVs, they can be used to infect a PEV physically and disseminate malware to other PEVs using cyber-physical links with EVCSs described in Section~\ref{cyber_physical}.

\subsubsection{ External Vulnerabilities}
Besides the attacks incurred via the CAN bus, PEVs face security threats from their back-end communication entities including
manufacturers, radio stations, road-side infrastructures, and other vehicles as shown in Fig \ref{fig:EV1}. The PEV vendors or OEMs send patches and data to the PEV wirelessly via cellular networks \cite{ECU_cellular,uptane}. In some vehicles patching is done manually using a USB drive. However, the wireless method is preferred over  manual patching due to its cost-effectiveness and swift delivery. On the other hand, this wireless data transmission opens a wide attack surface. For instance, an attacker can launch a man-in-the-middle attack on  cellular networks and inject malware into ECUs. Moreover, an attacker can launch various variants of DoS attacks such as dropping or delaying the patch requests, sending an older version of updates, and partly patching the ECUs \cite{uptane}. 

Similarly, the back-end connectivity of the PEV infotainment system to radio stations, road-side infrastructures (e.g., traffic signals), and other vehicles, which is the backbone of an emerging autonomous driving industry is vulnerable to attacks \cite{smith2016car}.  
Also, vulnerabilities have been exposed in the communication links between the EVCSs and PEVs. Thus, the signals exchanged between an EVCS and a PEV that carry  information about charging current can be spoofed, and, hence, over-or under-charge the battery pack and change the power consumed by PEVs from the power grid \cite{chpointcyber}. The short-range wireless communication channels such as Bluetooth, WiFi, and NFC in the PEV systems also expose attack surfaces (e.g., attacks on wirelessly operating vehicle door locks \cite{gracia2016lock}). 

\subsection{Vulnerabilities in the EVCS System}
\label{vul_EVCS}
The EVCSs can be compromised directly via on-site interactions or remotely through communication interfaces. The EVCS  cybersecurity analyses, such as in \cite{encs,DOE}, report widespread cyber vulnerabilities in the EVCS architecture. We categorize the vulnerabilities of EVCSs into  internal and external, which are relevant to provide a conceptual background to the demand-side cyberattacks in this paper. 

\subsubsection{Internal Vulnerabilities} 
An attacker with physical access to an EVCS can exploit the EVCS hardware and software via its peripherals such as USB ports \cite{inlEVCS,inlEVCS1,gottumukkala2019cyber}. Furthermore, this capability can be remotely exercised even if the attacker gets physical access only temporarily \cite{inlEVCS1}. The possibility of physical access depends upon factors such EVCS level, and tamper resistance of a particular EVCS location. In particular, public L3 EVCSs have greater physical exposure and physical access points (e.g., USB ports) than the residential L1 EVCSs. The physical entry points to the EVCSs, i.e., USB ports, serial ports, and Ethernet jacks are mounted outside the EVCS casing. Most of the EVCSs have processors running on a Linux kernel and communicate using RS232 protocols. These firmware reportedly may use  weak or default authentication credentials and message encryption technologies, which can be reverse-engineered \cite{inlEVCS, DOE, P3}. Furthermore, access control is weakly applied in the EVCS operating system, e.g., some processes not necessarily requiring root access are executed with root user privileges \cite{inlEVCS, Schneider}. Also,  EVCS processors use a shared memory configuration, which can allow for interfering with  the shared memory. The EVCS firmware extraction is possible using Joint Test Action Group, Asynchronous Receiver-Transmitter, flash memory readers and USB sticks\cite{gottumukkala2019cyber}. The extracted firmware can be leveraged to launch more sophisticated cyberattacks. 

Various attacks pertaining to the availability, confidentiality and integrity of EVCS services can be launched once the hardware and firmware are accessed. For instance, the EVCS charging command exchanged with a PEV can be eavesdropped, replayed or altered to launch an integrity attack on the EVCS service. Customers data on  authentication, billing, and charging history, which are stored locally in the EVCS, can be obtained, which would breach the privacy and confidentiality of PEV users. The attacker can stop the coordination between the EVCS controllers or turn off all power electronics modules of the EVCS and, hence, disrupt the EVCS operations and their remote control by the EVCS operator \cite{inlEVCS1}.

\subsubsection{External Vulnerabilities} 
Fig.~\ref{fig:EV1} shows various external communication interfaces of EVCSs. The complexity of these interfaces increases with an EVCS power level and so do their cyber vulnerabilities. 
The EVCS HMI interface uses either Radio Frequency Identification (RFID) tags or a smartphone app to authenticate PEV users. An attacker can reverse engineer the RFID tag and steal user information, gain unauthorized access to PEVs and EVCS, and even take down the EVCS \cite{RFID}. Further, the authentication of a PEV user in EVCS is done over cellular networks using a smartphone app, which increases the attack surface \cite{DOE,P3}. Also, the interface between an EVCS and an EVCS server and between an EVCS server and a PEV user is performed over cellular networks and a proprietary WAN technology. The attacker can compromise these interfaces by exploiting the vulnerabilities in cellular networks \cite{ECU_cellular}, the EVCS servers, and  smartphones of PEV users. The compromised EVCS server can deny to authenticate PEV charging sessions or can send false EVCS information (charging price and online status of the EVCS) to the PEV user leading to a DoS attack or changing power consumption of PEVs as needed for launching the attack. The EVCS interfaces with PEVs over the wired communication channel. Exploiting the many-to-many relationship of PEVs and EVCSs, an infected PEV can thus additionally compromise numerous EVCSs and PEVs. Moreover, the interface between the EVCS or EVCS server and the power grid operator or BEMS is another pathway an attacker can similarly exploit and obtain unauthorized access into the EVCS. Similar to the PEVs, the patches and software updates sent wirelessly by their  manufactures or OEMs to the EVCS are not authenticated.

\subsection{Attack Vector}
The attack vector of interest in this paper arises in an urban power grid environment, where a relatively high number of PEVs and EVCSs, as well as power consumption density, make it possible to  collect sufficient public data about the power grid, EVCSs and PEVs. This attack vector consists of PEVs with vulnerabilities described in Section~\ref{vul_PEV}, and L2 and L3 EVCSs, which have vulnerabilities described in Section~\ref{vul_EVCS}. For example, this attack vector can be realized as follows. The attacker can hack into EVCS servers (see Fig. \ref{fig:EV1}) using remote access networks, virtual private networks or wireless networks. After intrusion into the EVCS servers, the attacker can install malware that stealthily sends false charging commands to  PEV users as explained in Section~\ref{vul_EVCS}. Using the false charging commands, the attacker can simultaneously shutdown a pre-calculated number of EVCS loads that would result in a sudden demand decrease in the power grid, leading to an over-frequency event. As a result of the over-frequency event, over-frequency relays will trip as prescribed by the IEEE 1547 Standard, thus disconnecting frequency-sensitive equipment (large generators, substations) and causing load shedding. Regardless of the vulnerability exploited by the attacker, the demand-side, data-driven cyberattack described next can be executed using the attack strategy described in Section~\ref{attack_prep} and shown in Fig.~\ref{fig:attack_sequence}.

\vspace{5mm}
\section{Power Grid Model}
\label{model_conv}

Assuming that the attacker has the publicly accessible EVCS and the power grid data from Section \ref{data}, designing an attack strategy requires a power grid model that relates the data with the physics of the power grid operation. Malicious load alterations are anticipated to be small, relative to the total system demand, and swift so as not to alarm the system operator. The impact of these small disturbances on the power grid stability can be analyzed using the linearized stability theory \cite{kundur1994power, vittal}. The core assumption that underlies this theory is that small disturbances and the dynamic behavior of the power grid can be accurately modeled by linear power flow equations (e.g., DC approximation) and by  first-order ordinary differential equations (e.g., swing equation). The use of the DC power flow in this paper is justifiable because: (i) underground cables in Manhattan are short in length ($\leq 6.9$ miles, see Fig. \ref{fig:network}), which results in a small value of resistance,  and (ii) anticipated alterations in demand to launch an attack are smaller relative to the system load, which implies small voltage angle differences between connected nodes. This DC power flow assumption has been widely used in the power grid cybersecurity literature, e.g., \cite{li2015bilevel, soltan2018react, amini2018dynamic, huang2018cyber, soltan2016power, zhang2016physical}. The data collected and the grid model allow the attacker to seek a data-driven, load-altering action causing frequency instability in the power grid.

\subsection{Model}
\label{model}
We consider a power grid with $\mathcal N$ nodes with mutually exclusive subsets\footnote{If a node hosts both  the generator and load it can be split into two  nodes.} of generator nodes $\mathcal G \subseteq \mathcal N $ and load nodes $\mathcal L \subseteq \mathcal N $. Let $N = card(\mathcal N)$ be the number of nodes such that  $N=1+G+L$, which includes one slack (reference) bus, $G=card(\mathcal G)$, and $L=card(\mathcal L)$. Let $\delta_i$ and $\theta_j$ be the elements of vectors of nodal voltage angles at generator node $i \in \mathcal G$ and at load node $j \in \mathcal L$, respectively. The nominal (synchronous) angular speed $\omega_s= 2\pi f_s$, where $f_s =60~\text{Hz}$. 

Using the DC power flow approximation, we model the nodal power balance for generator and load nodes as:
\allowdisplaybreaks
\begin{subequations} \label{eq:dc_powerflow}
\begin{flalign}
P_{i}^{G} = &  \sum_{k \in \mathcal B}{Y_{ik}}\Delta \delta_i ,\quad \forall i \in \mathcal G,  \label{eq:gen} \\
P_{j}^{L} = & -\sum_{k \in \mathcal B}{Y_{jk}}\Delta \theta_j  ,\quad \forall j \in \mathcal  L, \label{eq:load}
\end{flalign}
where $Y_{ik}$ and $Y_{jk}$ are the imaginary parts of complex admittances between nodes $i$ and $k$  and nodes $j$ and $k$, respectively.
Further, $\Delta \delta_i$ and $\Delta \delta_j$ in 
\begin{flalign}
\Delta \delta_i =
    \begin{cases}
    \delta_i -\delta_k , \qquad \forall i \in \mathcal G,~ \forall k \in \mathcal G,  \\
    \delta_i -\theta_k , \qquad \forall i \in \mathcal G,~\forall k \in \mathcal L, 
    \end{cases} \\
\Delta \theta_j =
    \begin{cases}
    \theta_j -\delta_k , \qquad \forall j \in \mathcal L,~ \forall k \in \mathcal G, \\
    \theta_j -\theta_k , \qquad \forall j \in \mathcal L,~ \forall k \in \mathcal L.
    \end{cases}
\end{flalign}
\end{subequations}
 \textit{Remark 1 (The power flow model)}: The DC power flow given by $P=f(\theta,\delta)$ can be approximated as an AC power flow linearized around a given operating point given by $\{P,Q\}=f(\theta_0, \delta_0, V_0 )$, where subscript $0$ refers to the operating point, and $Q$ is the reactive power \cite{huang2018cyber,soltan2018react}. We used the DC power flow model to conservatively assess the least-possible scenario of the cyberattack, which can be leveraged as the worst-case scenario attack by the power grid operator to build defense schemes. However, the attack scheme presented in this paper also holds for the other power flow models.

In addition to the DC power flow model in Eq.~\eqref{eq:dc_powerflow}, we model the dynamic behavior of the power grid using the swing equation for every generator node $i \in \mathcal G$:
\begin{subequations} \label{eq:swing_all}
\begin{flalign}
M_{i}\dot{\omega_{i}} = & ~P^{M}_{i} - P^{G}_{i} - D^{G}_{i}\omega_{i},  \label{eq:swing_gen} \\
\dot{\delta_{i}} =&~\omega_i, \label{eq:del}
\end{flalign}
where $M_i$ and $D^G_i$ are the moment of inertia and damping coefficient of the generator at node $i$, $\omega_i$ is the angular speed difference between the speed of the rotor of the generator at node $i$ and the synchronous speed ($\omega_s$). $P^G_i$ is the electrical power output and  $P^M_i$ is the  mechanical power output of its turbine driving the generator at node $i$.

The balance between $P^G=\sum_{i \in \mathcal G}P^G_i$ and $P^L=\sum_{j \in \mathcal L}P^L_j$ determines the frequency stability of the power grid. The conventional controllable generators are set to maintain  $P^G \approx P^L$ by adjusting $P^M_i$ using in-feed of $\omega_i$ in real-time. This control action is enabled by the automatic generation control (AGC) system, which is composed of two parallel control loops with $\omega_i$ feedback: (i) a proportional gain (can be regarded as a droop control parameter), which reduces the frequency deviation swiftly, and (ii) an integral gain which slowly takes the frequency deviation to zero \cite{saadat1999power}. This controller is implemented as:
\begin{flalign}
P^M_{i} = -\bigg( K^P_i \omega_{i} + K^I_i \int_{0}^{T}\omega_{i}\bigg), \label{eq:pi_controller}
\end{flalign}
where $K^P_i$ and $K^I_i$ are pre-defined proportional and integral gain parameters, respectively, and the negative sign on the right-hand side of Eq. \eqref{eq:pi_controller} indicates that  $P^M_i$ is adjusted in the opposite direction to changes in $\omega_i$. The value of $K^P_i$ is set to reduce instant frequency excursions and the value of $K^I$ is set to  reduce frequency volatility over the time period $T$. 
\end{subequations}

Using the DC power flow in Eq.~\eqref{eq:dc_powerflow} and the dynamic model of generator in Eq.~\eqref{eq:swing_all}, we can model the power grid dynamics under disturbances. 
We use $P^M_i$ from Eq.~\eqref{eq:pi_controller} and $P^G_i$ from Eq.~\eqref{eq:gen} into Eq.~\eqref{eq:swing_gen} to obtain the resulting dynamics of generator nodes $i \in \mathcal{G}$ in Eq.~\eqref{eq:gen_final}. Similarly, for each load node $j\in \mathcal L$, we split nodal power in Eq.~\eqref{eq:load} to formulate Eq.~\eqref{eq:load_final}, where $\Delta P^L_j$ is the EVCS demand altered by the attacker at node $j$, and $\overline{P}^L_j$ and $D^{L}_j \theta_j$ are the compromised $j^{th}$ nodal loads. Note that $\overline{P}^L_j$ accounts for frequency-insensitive loads such as lights and $D^{L}_j \theta_j$ represents the frequency-sensitive loads such as HVAC units, $D^{L}_j$ is the load-damping coefficient at node $j$.  Thus, the system of equations representing the power grid dynamics is:
\allowdisplaybreaks
\begin{subequations} \label{eq:final_dynamic_model}
\begin{flalign}
\label{eq:gen_final}
\begin{split}
&M_{i}\dot{\omega_{i}}  = \!-\! \big(K^P_i \!-\!D^G_{i}\big) \omega_{i}\! -\!K^I\delta_i\! -\!\sum_{k \in \mathcal G}{Y_{ik}}(\delta_i -\delta_k) \\
&\qquad + \sum_{k \in \mathcal L}{Y_{ik}}(\delta_i -\theta_k), \quad \forall i \in \mathcal G,
\end{split}
\\
\label{eq:load_final}
\begin{split}
&0= \overline{P}^{L}_j - D^L_j \theta_j +\Delta P^{L}_j + \sum_{k \in \mathcal G}{Y_{jk}}(\theta_j -\delta_k) \\ 
& \qquad +\sum_{k \in \mathcal L}{Y_{jk}}(\theta_j -\theta_k) , \qquad \forall j \in \mathcal  L,
\end{split}
\\
\label{omega_final}
\dot{\delta_{i}}&=~\omega_i, \quad \forall i \in \mathcal{G}.
\end{flalign}
\end{subequations}
\allowdisplaybreaks[0]
The linearized dynamic power grid model in Eq.~\eqref{eq:final_dynamic_model} can be built by the attacker using public data described in Section~\ref{data}. The manipulated EVCS demand data is accounted by parameter $\Delta P^{L}_i$, while power grid parameters $M_i, D^G_i, D^L_i, Y_{ik},$ and $Y_{jk}$  and AGC parameters $K^P_i, K^I_i$ can be obtained using sources summarized in Table~\ref{tab:info_attack}. 
The AGC parameters, $K^P_i$ and $ K^I_i$, are dynamic and are not publicly released by power utilities. Therefore, the AGC parameters are typically adjusted manually by the attacker such that the original (pre-attack) dynamic power grid model is stable, whereas power grid parameters and EVCS demand data can be elicited using the method illustrated in Table \ref{tab:info_attack}.  
The dynamic model of the power grid in Eq. \eqref{eq:final_dynamic_model} can be represented as a linear time-invariant (LTI) state-space descriptor system:
\begin{subequations}
\begin{flalign}
E\dot{x} = \hat{A}x  + \hat{B}u, \label{statespace_des}
\end{flalign}
with the descriptor matrix $E \in \mathbb{R}^{(2G+L)\times (2G+L)}$, state matix $\hat{A} \in \mathbb{R}^{(2G+L)\times (2G+L)}$, control vector $\hat{B} \in \mathbb{R}^{(2G+L)\times 1}$ and state variable vector $x \in \mathbb{R}^{(2G+L)\times 1}$, as well as scalar input $u \in \mathbb{R}$. The descriptor system in Eq.~\eqref{statespace_des} is regularized as:
\begin{flalign}
\label{eq:statespace_reg}
\dot{x} = Ax + Bu,
\end{flalign}
where $A=E^{-1}\hat{A}$ and  $B=E^{-1}\hat{B}$. In terms of Eq.~\eqref{eq:final_dynamic_model}, state vector $x$ and control input $u$ are defined as:
\allowdisplaybreaks
\begin{flalign}
x &=[\delta,  ~\omega,  ~\theta]^T \label{eq:x} \\
u_j& =\Delta P_j^L + \overline P_j^L \label{eq:u}, \quad 
\end{flalign}
where $\delta \in \mathbb{R}^{G \times 1}$, $\omega \in \mathbb{R}^{G \times 1}$, $\theta \in \mathbb{R}^{L \times 1}$, $j$ is the attack launching node. In turn, matrices $A$ and $B$ are defined as:
\begin{flalign}
\label{eq:statespace_def}
A =& \overbrace{\begin{bmatrix}
I &\hspace{-2mm} 0& \hspace{-2mm}0\\
0&\hspace{-2mm} -M& \hspace{-2mm} 0 \\
0&\hspace{-2mm} 0& \hspace{-2mm} D^L\\
\end{bmatrix}^{-1}}^{E^{-1}}  \overbrace{\begin{bmatrix}
0&\hspace{-2mm}I&0\\
K^I \!+\!Y_{GG}& \hspace{-2mm} K^P\! +\!D^G& Y_{GL} \\
Y_{LG}&\hspace{-2mm} 0& Y_{LL}\\
\end{bmatrix}}^{\hat A}  \\
B= &\underbrace{\begin{bmatrix}
I &0&0\\
0&-M& 0 \\
0&0& D^L\\
\end{bmatrix}^{-1}}_{E^{-1}}  \underbrace{\begin{bmatrix}
0\\
0\\
\hat{I}\\
\end{bmatrix}}_{\hat B},
\end{flalign}
\end{subequations}
where $Y_{GG}\in \mathbb{R}^{G\times G}$, $Y_{GL} \in \mathbb{R}^{G\times L}$, $Y_{LG} \in \mathbb{R}^{L\times G}$ and $Y_{LL} \in \mathbb{R}^{L\times L}$ are submatrices of admittance matrix $Y = [ Y_{GG}~Y_{GL};Y_{LG}~Y_{LL} ]$. $M \in \mathbb{R}^{G\times G}$,
$D^G \in \mathbb{R}^{G\times G}$,
$K^P \in \mathbb{R}^{G\times G}$, $K^I \in \mathbb{R}^{G\times G}$, and $D^L \in \mathbb{R}^{L\times L}$ are the diagonal submatrices and $I^{G\times G}$ is an identity matrix. 
Vector  $\hat{I} \in \mathbb{R}^{L\times 1}$ has all elements set to zero except the nodes where the attack is launched, i.e., $\Delta P^L_j \neq 0$. Using the power grid model in Eq.~\eqref{eq:statespace_reg}, the power grid stability can be evaluated using eigenvalues of state matrix $A$. These eigenvalues are the roots of the characteristic equation of the dynamic system in Eq.~\eqref{eq:statespace_reg}. The eigenvalues in the complex plane corresponds to the time-domain response of $x$. The exact relationship between eigenvalues and state variables can be computed using participation factors \cite{kundur1994power}. The attacker can use Eq.~\eqref{eq:statespace_reg} to estimate the eigenvalues of $A$ without malicious load alterations, i.e.,  $\Delta P^L =0$, and use this information  to seek $\Delta P^L \neq 0$ that modifies eigenvalues to cause instability.

\subsection{Data-driven Demand-Side Cyberattack}
\label{attack_data_driven}
After obtaining the data-driven power grid model, an attacker can design a data-driven demand-side cyberattack using the public EVCS data. The sophistication of the attack design is idiosyncratic and depends on various factors such as the type of the attacker, i.e., state, non-state, or individual actors. Since, this paper aims to emphasize the threat of leveraging the public data by an attacker irrespective of their type, we design a full state-feedback based demand-side cyberattack model. This design can inform the attacker on minimum requirements for the attack to succeed (e.g., calculating the required EVCS demand, identifying a relatively weak area in the power grid, and the most impactful time for the attack).

\subsubsection{Data-Driven Attack Mechanism} 
\begin{figure}[b]
\centering
\includegraphics[width=1.0\columnwidth, clip =true,trim = 10mm 0mm 0mm 0mm]{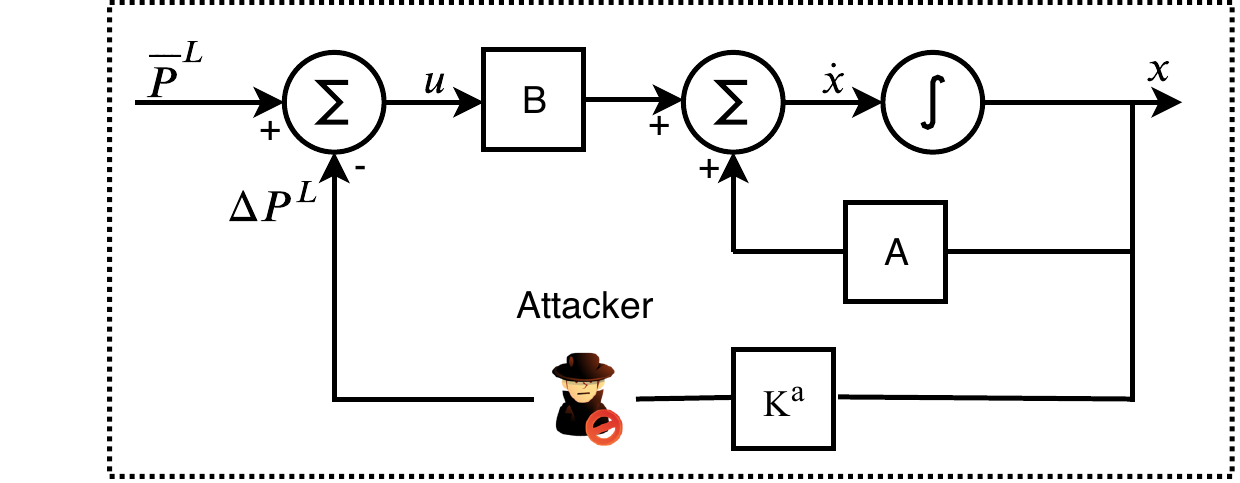}
\caption{Schematic representation of the demand-side cyberattack, where an attacker calculates the amount of compromised loads by adjusting $K^a$.}
\label{fig: ss_feedback}
\end{figure}

To design the attack,  we consider $u = \overline{P}^{L} + \Delta P^L = \overline{P}^{L} - K^a x,$  where $K^a \in \mathbb{R}^{1\times 2G+L} $ is the vector of gains   set by the attacker, which is proportional to the amount of the manipulated EVCSs demand. Hence,  the system in Eq.~\eqref{eq:statespace_reg} is recast as follows:
\begin{subequations} \label{eq:attack}
\begin{flalign}
\label{eq:feedback}
\dot{x}& = Ax +B(\overline{P}^{L}-K^ax) =(A-BK^a)x +B\overline{P}^{L},
\end{flalign}
which corresponds to the state-feedback based control diagram  in Fig.~\ref{fig: ss_feedback}. The stability of the system in Eq.~\eqref{eq:feedback} is determined by eigenvalues of matrix $(A-BK^a)$ and, thus, can be influenced by the attacker by strategically selecting the values of $K^a$. In turn, the attacker is limited in their ability to select the value of $K^a$ by the EVCS demand availability:
\begin{equation}
\begin{split}
0 \leq |K^ax| \leq \Delta P^{L, \max}, 
\end{split}
 \end{equation} 
 \end{subequations}
 where $\Delta P^{L, \max}$ is the maximum capacity of the EVCS demand that can be compromised. The main difficulty in implementing such dynamic attacks is the calculation of state vector $x$, which continuously changes over time. For example, the attack based on real-time measurements and feedback of $\omega_i \in x$ is presented in \cite{amini2018dynamic, pasqualetti2019}. However, for the preparation phase of the attack (see Section~\ref{attack_prep}), we use $x =[\delta,  ~\omega,  ~\theta]^T$ in the feedback to account for the dynamics of the whole power grid. The inclusion of the dynamics will result in a more accurate calculation of $\Delta P^L$. This inclusion is possible because of the data-driven power grid model developed in Sections \ref{data} and \ref{model}. Even if some of the parameters of the power grid model change over time, i.e., it directly affects matrix $A$ and vector $x$, the attacker can be informed of these changes via a public disclosure process of power utilities (e.g., real-time announcement about outages, maintenance, planned asset retirement/installations, upgrades, generation and demand schedules). Furthermore, topological modifications of the power grids are rare and manual. Hence, the attacker can track changes in power grid topology and operations, and modify their grid model and the attack. This capability allows remote attackers to prepare the attack without needing real-time measurements of the power grid states.
 
\vspace{2mm}
\subsubsection{Data-Driven Attack Optimization}  
Although the attacker can modify the eigenvalues of the power grid model to cause instability \cite{amini2018dynamic}, this relocation might be tracked by the grid operator. Thus, the relocation must be carried out in such a way that load alterations are kept to a  minimum. The attacker will aim to minimize $K^a$ to avoid being detected. Since the value of $K^a$ should be large enough to cause instability (e.g., to make the real part of at least one eigenvalue greater than or equal to zero), the attacker  faces an optimization problem of selecting the least-possible value of $K^a$ that ensures instability. To do this, we use the Bass-Gura approach for the state-feedback-based partial eigenvalue placement \cite{kailath1980linear}.

Let  $o(s)$ and $p(s)$ be the monic characteristic equations of the original (pre-attack) and compromised power grid models in Eqs.~\eqref{eq:statespace_reg} and \eqref{eq:feedback}, respectively. We obtain the eigenvalues by solving:
\allowdisplaybreaks
\begin{flalign}
       &o(s) \!=\!|sI\!-\!A| = s^n \!+\!o_{n-1}s^{n-1} \!+\! \ldots\!+\!o_{0}s^{0} =0 , \label{eq:Os} \\
        & p(s)\!=\!|sI\!-\!A\!+\!BK^a|\! =\! s^n \!+\!p_{n-1}s^{n-1} \!+\! \!\ldots\! \!+\!   p_{0}s^{0}\!=\! 0, \label{eq:Ps}
\end{flalign}
where $n= 2G + L$ is the order of the system and  $e^o \in \mathbb{C}^{n\times1}$ and  $e^p \in \mathbb{C}^{n\times1}$ are vectors of eigenvalues for Eqs.~\eqref{eq:Os}-\eqref{eq:Ps}.  We assume that the original power grid model is stable, i.e., $\Re(e^o)<0$. Adjusting $K^a$ the attacker modifies eigenvalues in $e^o$ such that some eigenvalues in $e^p$  become real positive. Using the state-feedback based controller design procedure for a fully controllable system, presented in \cite{kailath1980linear,6144706}, the relationship between the coefficients of Eq.~\eqref{eq:Os} and those of Eq.~\eqref{eq:Ps} can be written using $n$ linear equations in terms of $K^a$. In the matrix form, the equations can be written as:
\begin{equation}
\label{eq:kali}
    p -o = {W^T}{M}_cK^a,
\end{equation}
where $p = [p_{0} ~p_{1} \ldots p_{n-1}]^T$ and $o= [o_{0} ~o_1 \ldots o_{n-1}]^T$ are vectors of coefficients of the characteristic equations of the original and compromised power grid model, respectively, ${W} \in \mathbb{R}^{n\times n}$ is a Hankel matrix
with its first column set to $[o_{1}~ o_{2}\ldots o_{n-1}~1]^T$ and elements below the  anti-diagonal are zero, and ${M}_c  \in \mathbb{R}^{n\times n}$ is a controllability matrix defined as:
\begin{equation}
\label{eq:controllability}
{M}_c = [ B ~ AB \ldots A^{n-1}B],
\end{equation}
where  $rank({M}_c)$ defines the maximum number of eigenvalues that can be relocated. For instance, if $rank({M}_c)<n$, the system is partially controllable for a given input, i.e., only $rank({M}_c)$ eigenvalues can be relocated arbitrarily on the complex plane by changing the EVCS demand at a given node.
Since matrix ${M}_c$ is composed of matrices $A$ and $B$, its rank and the number of eigenvalues that can be relocated depends on parameters $M$, $D^G$, $Y$, $K^P$, $K^I$, and $D^L$, which can be learned by the attacker using public sources. 

The attacker can determine the value of $K^a$ using Eq.~\eqref{eq:kali} for some given vector $p$ because vector $o$ and matrices $W^T$ and $M_c$ are computed using parameters in Eq.\eqref{eq:statespace_reg}. The attacker defines $p$ in such a way that at least one eigenvalue in $e^p$ becomes real positive. The maximum number of the real positive eigenvalues that an attacker can choose is upper bounded by $rank({M}_c)$, because only $rank({M}_c)$ eigenvalues can be arbitrarily relocated on the complex plane.
Recasting $p(s)$ in Eq.~\eqref{eq:Ps} in the decomposed polynomial form in terms of eigenvalues leads to  \cite{6144706}:
\begin{equation}
\label{eq:decompose}
p(s) =\underbrace{\prod_{i=1}^{m} (s+e^a_i)}_{a(s)} \underbrace{\prod_{j=1}^{n-m} (s+e^r_j)}_{r(s)},
\end{equation}
where $m\leq rank({M}_c)$ is the number of eigenvalues that the attacker attempts to relocate, $e^a \subseteq e^p$ is the vector  of eigenvalues defined by the attacker for relocation and  $e^r \subseteq e^p$ is the vector of the remaining eigenvalues in $e^p$ such that $card(e^a)+card(e^r)=card(e^p)$. Vectors $a = [a_{0}~ a_{1} \ldots a_{m-1}]^T$ and $r = [r_{0}~r_{1}\ldots r_{n-m-1}]^T$ are  coefficients of monic polynomials formed by $e^a$ and $e^r$, respectively. Given $e^a$, we use Eqs.~\eqref{eq:kali} and \eqref{eq:decompose} to compute $r$ \cite{6144706}:
\begin{flalign}
r = F{W^T}{M}_cK^a + g, \label{eq:es}
\end{flalign}
where $F \in \mathbb{C} ^{n-m+1 \times n}$ and $g \in \mathbb{C} ^{n-m+1 \times 1} $ are the auxiliary matrix and vector, respectively, defined by Eqs.~\eqref{eq:F1}--\eqref{eq:gend}. Since the matrices ${M}_c$ and ${W}$ are derived from matrices $A$ and $B$, the auxiliary terms are fully parameterized as:
\allowdisplaybreaks
\begin{subequations}
\begin{flalign}
\label{eq:F1}
&F(i,1) = \sum_{k=1}^{i-1} F(i-k,1) \frac{-a_{k}}{a_0}, \quad   \forall i = 2\ldots n-m, \\ 
&F(1,1) = \frac{1}{a_0}, \quad F(1,j) =0 , \quad  \forall j =2\ldots n,\\
&F(i,j) = F(i-1,j-1),\forall i =2\ldots n-m, \forall j =2\ldots m, \\
\label{eq:Fend}
&F(n-m+1,j) =0, \quad \forall j =2\ldots m, \\\label{eq:g1}
&g(i) =\sum_{k=1}^{i-1} (g(i-k)\frac{-a_{k}}{a_0}) +\frac{o_{i-1}}{a_0}, \forall i = 2\ldots n-m, \\
\label{eq:gend}
&g(1) = \frac{o_0}{a_0}, \quad g(n-m+1) =1.
\end{flalign}
\end{subequations}
\allowdisplaybreaks[0]
Back substituting vector $r$ in Eq.~\eqref{eq:decompose} returns an expression for vector $p$, which if substituted in Eq.~\eqref{eq:kali} yields $m$ linearly independent nonzero equations in terms of $K^a$: 
\begin{equation}
V{W^T}{M}_cK^a +h =0, \label{eq:final_form}
\end{equation}
where matrix $V \in \mathbb{C}^{m \times n}$ and vector $h\in \mathbb{C}^{m\times 1}$ are parameterized in terms of matrices $A$ and $B$, and assures $e^a \subseteq e^p$ as desired by the attacker.  Since Eq.~\eqref{eq:final_form} governs the eigenvalue relocation, the attacker can use the following optimization problem to select the least possible value of $K^a$:
\begin{subequations}\label{eq:opt}
\begin{flalign}
  &\underset{{K^a} \in \mathbb{R}^n}{\min} ||K^a||_2 \\
  &V{W^T}{M}_cK^a +h =0, \\
  &0 \leq |K^ax| \leq \Delta P^{L, \max}. \qquad  \label{eq:opt1} 
\end{flalign}
\end{subequations}
\subsubsection{Parameter Uncertainty in the Data-Driven Attack} 
\label{chance_constarint}To account for the likelihood of erroneous EVCS data, the attacker may robustify the data-driven attack optimized in Eq.~\eqref{eq:opt} against inaccuracies of the model parameters it learns. Randomness in $\Delta P^{L, \max}$ can be modeled as $\Delta P^L(\epsilon) =\Delta P^{L, \max} +\epsilon$, where $\epsilon$ is the model parameter uncertainty or inaccuracy (e.g.,  Gaussian noise). Thus, Eq.~\eqref{eq:opt1} is replaced with the following probabilistic constraint:
\begin{equation}
    \mathbb{P} \big(|K^ax| \leq \Delta P^L(\epsilon)\big) \geq 1-\eta, \label{eq:cc}
\end{equation}
where $\eta$ is a small number chosen by the attacker based on their confidence in the data. Eq.~\eqref{eq:cc} can be reformulated as a second-order conic constraint \cite{lobo1998applications}: 
\allowdisplaybreaks
\begin{flalign}
    -\Delta P^{L\max} + \alpha \leq K^ax \leq \Delta P^{L\max} -\alpha,
\end{flalign}

\allowdisplaybreaks[0]
where $\alpha=\phi^{-1}(\eta)\text{Stdev}(\epsilon)$ is an error margin on estimating $\Delta P^{L\max}$ and $\phi^{-1}$ is  an inverse cumulative
distribution function of the standard Gaussian distribution with zero mean. 

\section{Case Study}

We evaluate the feasibility of the data-driven attack developed above using the EVCS demand and power grid data illustrated in Fig.~\ref{fig:network_man}. 
The expected value and standard deviation of the EVCS demand is used as in Fig.~\ref{fig:weekall} with $\eta = 0.005$.
We set node B7, connecting Manhattan, NY with New Jersey (see Fig.~\ref{fig:network_man}), as the reference node since it is the largest power supplier to Manhattan, NY. The base power is 100 MVA and the rated system frequency is 60 Hz.  
The values of  state vector $x$ in the optimization problem in Eq.~\eqref{eq:opt} are conservatively obtained by solving Eq.~\eqref{eq:statespace_reg} for the operating condition pertaining to the tripping of the generator at node B7. The generator tripping is assumed to occur when the frequency exceeds 62 Hz for more than 0.16 seconds, in accordance to the IEEE Standard 1547.

\textit{Remark 2 (The value of $x$)}: State vector $x$ is dynamic and its value differs with the power grid operating conditions, i.e., the change in 
matrix $A$ and vector $B$. Hence, the attacker need to evaluate the required amount of the EVCS demand to be manipulated (i.e., $\Delta P^L=K^ax$) in a case-specific manner. However, this evaluation is computationally affordable because: (i) changes in matrix $A$ are trackable, and (ii)  vector $B$ varies with the interested node of attack. 

The case study uses the CVX package run under MATLAB and is carried out on a MacBook Air with a 2.2 GHz Intel Core i7 processor and 8 GB RAM. The optimization in Eq.~\eqref{eq:opt} is convex and, hence, does not pose a computational challenge, even if applied to larger networks  than in Fig.~\ref{fig:network_man}. All instances below were solved under tens of seconds. 

\subsection{Ability of the Attacker to Relocate Eigenvalues}
\begin{figure}[t]
\centering
\includegraphics[width=0.95\columnwidth, clip =true, clip =true, trim = 10mm 0mm 0mm 5mm]{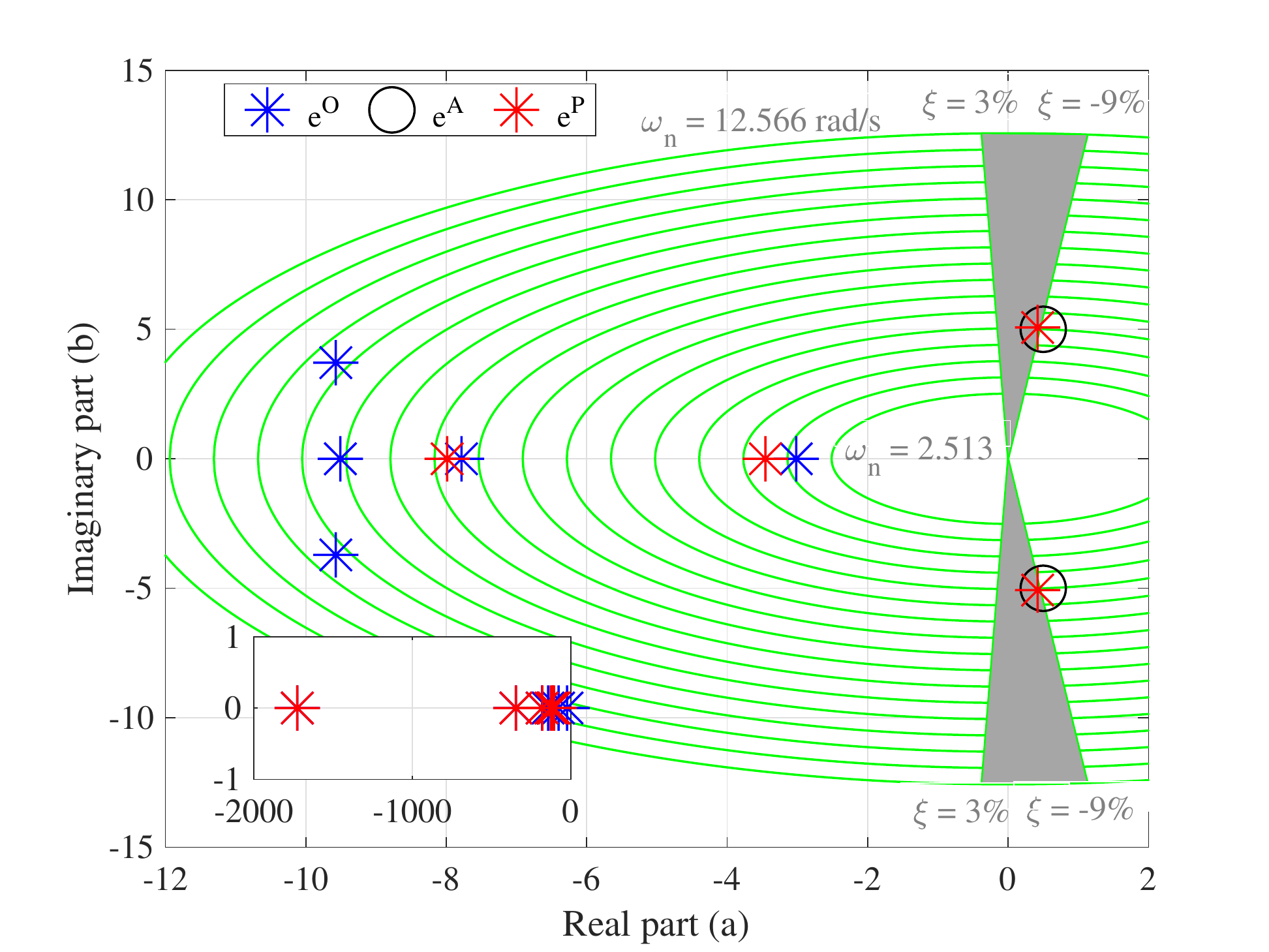}
\caption{Relocation of the eigenvalues under attack on node B4, where $e^o$ denotes original (pre-attack) eigenvalues and $e^a$ denotes  eigenvalue locations targeted by the attacker. The post-attack eigenvalues  are denoted as $e^p$.  Green lines represent $\xi$ and $\omega_n$ and the gray shaded area represents $S^a$. }
\label{fig:poles}
\end{figure}

The objective of this subsection is to demonstrate that the attacker can leverage the optimization in Eq.~\eqref{eq:opt} to move eigenvalues of the pre-attack system to pre-determined locations in the real-positive plane, thus causing system instability. The power grid in Fig.~\ref{fig:network} has 4 generation nodes (including  reference node B7) represented by $(\omega_i, \delta_i)$, and 4 load nodes represented by $\theta_j$. Thus,  state vector $x$ has 12 entries (2 entries per generation node and 1 entry per load node) and the pre-attack power grid modeled by Eq.~\eqref{eq:statespace_reg} has 12 eigenvalues. We select node B4 to attack because it has the greatest demand. Using Eq. \eqref{eq:controllability}, we compute  controllability matrix ${M}_c$  and find that  $rank({M}_c) =2$, which means that the attacker can attempt to move up to 2 eigenvalues. Since the objective of the attacker is to move these eigenvalues to the  real-positive plane to destabilize the power grid, the target eigenvalue locations are arbitrarily set to $e^a = a \pm jb=  0.5\pm j5$ for the demonstration as shown in Fig.~\ref{fig:poles}. The eigenvalues can be represented in terms of damping ratio $\xi$ and natural oscillation frequency $\omega_n$ as $a =-\xi \omega_n$ and $b=\omega_n \sqrt{1-\xi^2}$ \cite{kundur1994power}. Taken together, parameters $\xi$ and $\omega_n$ characterize the time-domain response (in terms of decay or increase in the amplitude of oscillations) of state vector $x$. Hence, $e^a = a \pm jb=  0.5\pm j5$ corresponds to the damping ratio of  $\xi =-10\%$ and the natural oscillation frequency of $\omega_n =5~rad/s$. Upon relocating  2 eigenvalues to the  target locations in the real-positive plane, 2 out of 12 state variables will oscillate following the attack with angular frequency $\omega_n$ and an increasing amplitude (due to  negative damping $-\xi$),  causing  power grid instability.

 Under the attack scenario described above, the current maximum EVCS demand given by  the maximum daily peak of $\approx$600 kW and standard deviation of 211 kW (both are observed at 14:00, see Fig. \ref{fig:weekall}) is not sufficient to relocate any eigenvalue to the target locations and, therefore,  Eq.~\eqref{eq:opt} yields an infeasible solution. Therefore, the EVCS demand at node B4 is scaled up to the maximum daily peak of 355 MW\footnote{Equivalent to $\approx$2,900 Model S Teslas simultaneously charged by 120 kW superchargers. The number  reduces to $\approx$1000 PEVs, if 350 kW Ionity high-power chargers are used instead.} and standard deviation of 124 MW to simulate a higher PEV penetration case. The increase in the EVCS demand to 355 MW is such that its simultaneous manipulation excurs the power grid frequency to 62 Hz for more than 0.16 seconds. In this case, two eigenvalues are moved into the real-positive plane as shown in Fig.~\ref{fig:poles}, which causes power grid instability targeted by the attacker.

\subsection{Minimum EVCS Demand to Destabilize the Power Grid}
\label{min_EVCS}
In the previous subsection, we demonstrate the feasibility of the data-driven attack for the case with a relatively high, but foreseeable  penetration rate of PEVs. However, this demonstration used arbitrary target eigenvalue locations. In practice, it is anticipated that relocation of eigenvalues can be detected by the power grid operator. Therefore, the attacker is likely to mask its intention and relocate eigenvalues surreptitiously. In this scenario, the attacker may elect to move eigenvalues to a region of vulnerability, where endogenous disturbances natural to power grid operations can cause power grid instability. The North American Electric Reliability Corporation defines the region of vulnerability as, \cite{tabrizi2017power}:
\begin{equation}
  S^a \in \mathbb{C}: \{\xi \leq 3\%, \quad 2.5\leq \omega_n \geq 12.6~rad/s\}.  \label{eq:region_of_vulnerability}
\end{equation}
Using the region of vulnerability in Eq.~\eqref{eq:region_of_vulnerability}, we will compute the minimum EVCS demand that the attacker needs to compromise to move 2 eigenvalues into that region of vulnerability. We discretize $S^a$ using a resolution of 0.3\% and $0.1~rad/s$ intervals for $\xi$ and $\omega_n$, respectively, and obtain the discrete space $\hat S^a$. For each pair $\{\hat \xi, \hat \omega_n \} \in \hat S^a$, we compute target eigenvalues as $\hat e^a = \hat a \pm j \hat b$, where $\hat a=-\hat \xi \hat \omega_n$ and $\hat b=\hat \omega_n \sqrt{1-\hat \xi^2}$. In addition to the selection of target eigenvalues $\hat e^a$ that the attacker seeks to achieve, the amount of EVCS loads that the attacker needs to compromise in order to launch an attack depends on pre-attack eigenvalues $e^o$ and $rank({M}_c)$. 
Since $rank({M}_c) < card (x)$, the attacker can directly impact\footnote{The state variables directly impacted by the attack can be inferred using the participation factor \cite{kundur1994power}, which defines  relationships between eigenvalues $e^o$ and state variables $x$, and  relative locations of $e^o$ and $e^p$. For instance, the attack in Fig. \ref{fig:poles} moved $e^o =-9.6\pm j0.0037$ to $e^a =0.5\pm j5$, which directly impacts state variable $\delta$ of nodes B4 and B7, and state variable $\theta$ of nodes B4 and B5, while the remaining state variables in $x$ are barely affected.} only some state variables in $x$. As a result, the attacker cannot always relocate eigenvalues to the chosen target locations precisely. Therefore, for each pair $\{ \hat \xi, \hat \omega_n \} \in \hat S^a$, we obtain the value of $\hat K^a$ using the optimization in Eq. \eqref{eq:opt} and obtain the minimum EVCS load ($\Delta \hat P^L=\hat K^ax)$ that needs to be compromised to destabilize the power grid. To assess how precisely the attacker managed to relocate eigenvalues to the target locations, we use  distance metric $\varepsilon = ||\tilde e^p-\hat e^a||_2$, where $\tilde e^p \subseteq e^p$ is the vector of the 2 nearest eigenvalues to $\hat e^a$. This distance serves as a measure of remoteness between the actual position of the 2 nearest and  target eigenvalue locations. 

\begin{figure}[!t]
\centering
\includegraphics[width=0.88\columnwidth, clip =true,trim = 0mm 0mm 0mm 0mm]{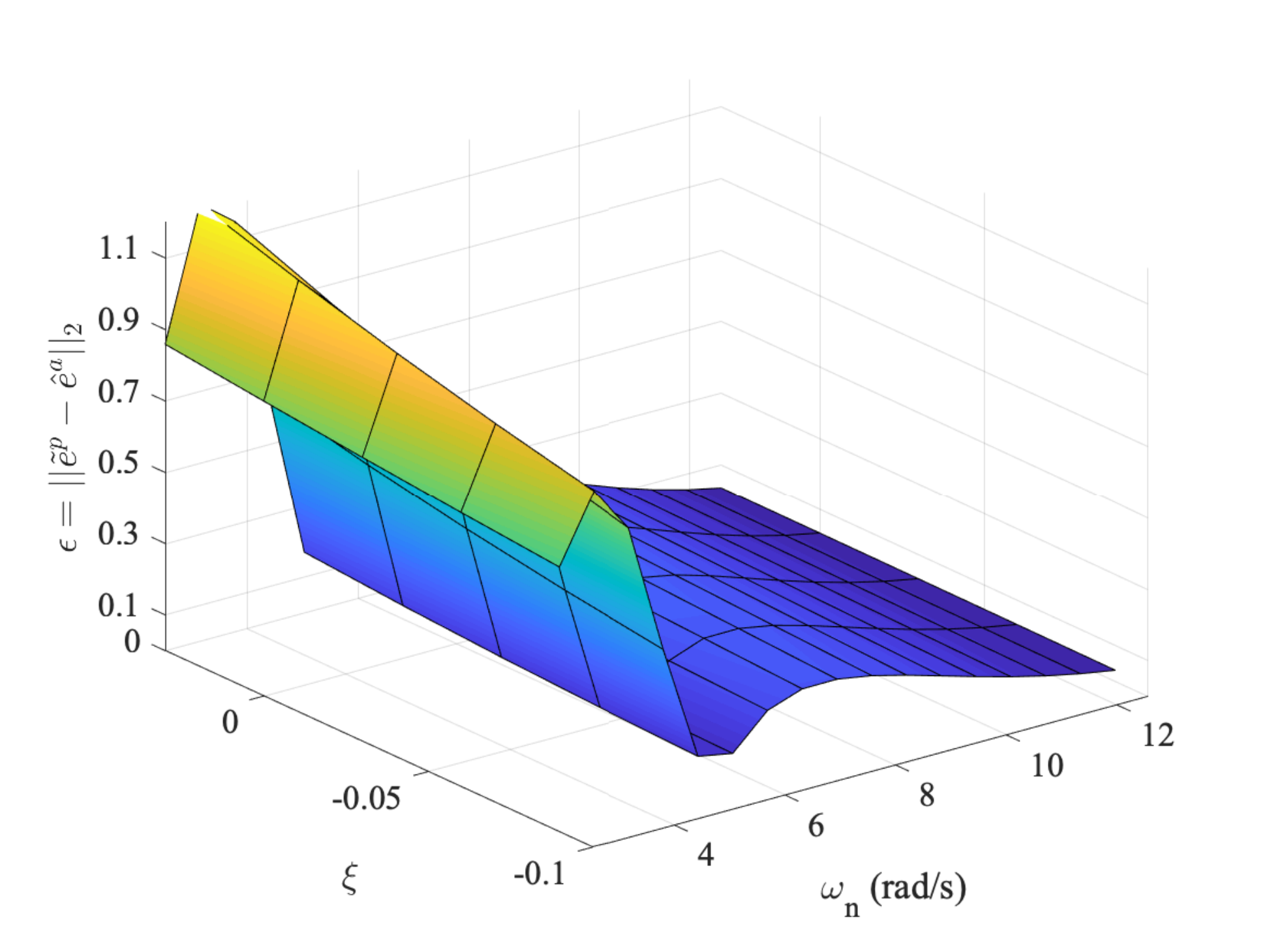}
\caption{Maximum relocation error $\varepsilon = ||\tilde e^p-\hat e^a||_2$ for different $\{ \hat \xi, \hat \omega_n \} \in \hat S^a$ chosen by the attacker, where  $\tilde e^p$ are the two nearest eigenvalues  to $\hat e^a$. }
\label{fig:error}
\end{figure}

\begin{table}
\caption{$\Delta \hat  P^L$ Needed to Move  Eigenvalues with the Relocation Error of $\varepsilon \leq 0.1$ (MW)}
\centering
\begin{threeparttable}
    \begin{tabular}{c|c|c|c|c|c}
    \hline
    \hline
         \multirow{2}{*}{$\omega_n( rad/s)$}&\multicolumn{4}{c}{$\xi$}\\
    \cline{2-6} &-0.09&-0.06&-0.03&0&0.03\\
    \hline
    5.7 &341.5&336.3&331.6&326.7&321.8\\
    10.7 &290.5&283.2&N/A\tnote{*}&N/A\tnote{*}&N/A\tnote{*}\\  
    11.3&282.9&275.8&268.8&261.7&254.7 \\
    11.9&275.1&268.3&261.6&254.9&248.3 \\
    12.6&267&260.5&254.2&247.9&241.6\\
    \hline \hline
    \end{tabular}
\begin{tablenotes}\footnotesize
\item[*]Value corresponds to $\varepsilon > 0.1$ and labeled not available (N/A)
\end{tablenotes}
\end{threeparttable}
\label{tab:feasiblePL}
\end{table}

Fig.~\ref{fig:error} illustrates the ability of the attacker to relocate eigenvalues to the target locations precisely. The relocation accuracy improves, i.e., the  value of $\varepsilon \rightarrow 0$, as  the values of $\hat \omega_n \in \hat{\mathcal{S}}^a$ and $\hat \xi \in \hat{\mathcal{S}}^a$ increase. However, the value of $\varepsilon$  is more sensitive to  $\hat \omega_n$ than to $\hat \xi$. The attack scenarios that use pairs $\{ \hat \xi, \hat \omega_n \} \in \hat S^a$ with a relatively high accuracy of relocation (e.g., $\varepsilon < 0.1$, see Fig.~\ref{fig:error}) are used to compute $\Delta \hat P^L = \hat K^ax$ and are summarized in Table~\ref{tab:feasiblePL}. As the value of $\xi$ and $\omega_n$ increase, the minimum EVCS load required to launch an attack from node B4 reduces. In other words, the amount of EVCS loads needed to be compromised increases with the severity of instabilities, i.e., more EVCS demand is required to be manipulated for higher oscillations and negative damping of the time-domain response of  state variables $x$. 

This analysis assumes that the data-driven attack is launched from node B4. If this case study is carried out for other nodes, the severity of the attack reduces. For example, if the attack is launched by nodes B3, B5, and B6, it will not destabilize the power grid as their load is not enough to move the eigenvalues, with smaller amount of $\varepsilon$, into the region of vulnerability.

\subsection{Sensitivity of the  Data-driven Attack Model} 
 
The demand-side attack model has power grid parameters $M, D^G, D^L,$ and $Y$, AGC parameters $K^P$ and $K^I$, and EVCS demand $\Delta P^{L}$. Among these, the AGC parameters are adjusted manually by the attacker such that the open-loop power grid model in Eq.~\eqref{eq:statespace_def} is stable, i.e., all the eigenvalues of matrix $A$ are in the left half of complex plane or left of the region of vulnerability $S^a$ defined in Eq.~\eqref{eq:region_of_vulnerability}. However, the power grid parameters are determined using public data as explained in Table~\ref{tab:info_attack}. Table~\ref{tab:sensitivity} analyzes the effect of erroneous or inaccurate power grid data on the demand-side attack in terms of the minimum EVCS demand required to destabilize the power grid, and the corresponding point in the region of vulnerability ($\xi$ and $\omega)$. The error in estimated power grid data is introduced uniformly across all power grid parameters.  In all the cases, except for $+5\%$ and $-5\%$ error, the attack relocated the eigenvalues to the same location as in the case with no error (see Section~\ref{min_EVCS}) in $S^a$ with relocation error $\varepsilon$ $\leq 10\%$. Furthermore,  since the relocation of eigenvalues is not linear, a greater EVCS demand is essential to destabilize the power grid, if the attacker acquired erroneous or inaccurate power grid data. If there is a $100\%$ error in the power grid parameters, the demand-side cyberattack is not feasible. 

\begin{table}
\centering

\caption{$\Delta \hat  P^L$ Needed to Move  Eigenvalues in the Region of Vulnerability with the Relocation Error of $\varepsilon \leq 0.1$ (MW)}
\centering
\begin{threeparttable}
    \begin{tabular}{c|c|c|c}
    \hline
    \hline
         \makecell{\% Error in \\grid parameters}& $\Delta \hat  P^L$ (MW) &$\xi$ & $\omega_n( rad/s)$\\
    \hline
    -50& 278.1&0.03&12.566\\
    \hline
    -10&262.1&0.03&12.566\\
    \hline
    -7.5&261.07&0.03&12.566\\
    \hline
    -5&260.8&0.03&12.566\\
    \hline
    -2.5&260.6&0.03&12.566\\
    \hline
    +2.5& 261&0.03&12.566\\
    \hline
    +5&334.5&-0.03&6.2832\\
    \hline
    +7.5&260&0.03&12.566\\
    \hline
    +10&260&0.03&12.566\\
    \hline
    +50&260.7&0.03&12.566\\
    \hline
\end{tabular}
\end{threeparttable}
\label{tab:sensitivity}

\end{table}

\section{Conclusion and Future Work}
This paper unveils a demand-side cyberattack that can imperil the power grid operations using PEVs and EVCS infrastructure. The attack uses publicly available EVCS and power grid data to design a data-driven attack strategy to destabilize the power grid using partial eigenvalue relocation. Using data-driven optimization, we study the impact of this attack on the power grid of Manhattan, NY. Although the current PEV penetration does not seem feasible to hamper the power grid stability, it highlights an emerging vulnerability as more PEVs are rolled out. In the future, there will be more  high-wattage EVCSs and  more PEVs charging simultaneously.

Due to the complicated nature of demand-side cyberattacks, which differs from direct cyberattacks on the utility-end assets, it is important to create new ways of detecting and mitigating them, which must accommodate limited, if any, observability of demand-side high-wattage appliances by utilities and demand-side restrictions (e.g., privacy of PEV users). We explore this important extension in our ongoing research work. Furthermore, we will extend the current attack vector, exploiting EVCS and PEV vulnerabilities, to include other high-wattage demand-side appliances (e.g.,  air-conditioners, heat pumps, boilers) and their unique characteristics. Finally, it is important to improve the accuracy of the power grid model built using publicly available data. Future extensions may include power grid non-linearities (e.g., AC power flows, saturation effect of conventional generators, inverter-based renewable energy resources) in the power grid model.

\bibliographystyle{IEEEtran}
\bibliography{references}
\end{document}